\documentclass[a4paper]{article}

\usepackage{graphicx,fullpage,placeins}
\usepackage[sort&compress,authoryear,semicolon,square]{natbib}
\renewcommand{\cite}{\citep}

\graphicspath{{.}{figures/}}
\usepackage{algorithm,algpseudocode,amsmath,url,gnuplottex} 

\usepackage[dvipsnames]{xcolor}
\usepackage{listings}

\newcommand\YAMLcolonstyle{\color{red}\mdseries}
\newcommand\YAMLkeystyle{\color{black}\bfseries}
\newcommand\YAMLvaluestyle{\color{blue}\mdseries}

\makeatletter

\newcommand\language@yaml{yaml}

\expandafter\expandafter\expandafter\lstdefinelanguage
\expandafter{\language@yaml}
{
  keywords={true,false,null,y,n},
  keywordstyle=\color{darkgray}\bfseries,
  basicstyle=\YAMLkeystyle,                                 
  sensitive=false,
  comment=[l]{\#},
  morecomment=[s]{/*}{*/},
  commentstyle=\color{purple}\ttfamily,
  stringstyle=\YAMLvaluestyle\ttfamily,
  moredelim=[l][\color{orange}]{\&},
  moredelim=[l][\color{magenta}]{*},
  moredelim=**[il][\YAMLcolonstyle{:}\YAMLvaluestyle]{:},   
  moredelim=**[il][\YAMLcolonstyle{ }\YAMLvaluestyle]{;},   
  morestring=[b]',
  morestring=[b]",
  literate =    {---}{{\ProcessThreeDashes}}3
                {>}{{\textcolor{red}\textgreater}}1     
                {|}{{\textcolor{red}\textbar}}1 
                {\ -\ }{{\mdseries\ -\ }}3,
}

\lst@AddToHook{EveryLine}{\ifx\lst@language\language@yaml\YAMLkeystyle\fi}
\makeatother

\newcommand\ProcessThreeDashes{\llap{\color{cyan}\mdseries-{-}-}}

\usepackage{silence}
\usepackage[font=footnotesize, labelfont=sf]{caption}
\usepackage[font=footnotesize, labelfont=sf]{subcaption}
\makeatletter
\renewcommand{\ALG@name}{ALGORITHM}
 
\begin{document}  

\title{Context Trees: Augmenting Geospatial Trajectories with Context}
\author{Alasdair Thomason, Nathan Griffiths, Victor Sanchez\\
~\\
Department of Computer Science, \\University of Warwick, UK}
\date{June 2016}

\maketitle

\let\svthefootnote\thefootnote
\let\thefootnote\relax\footnote{Authors' contact details: \{Alasdair.Thomason, Nathan.Griffiths, V.F.Sanchez-Sliva\}@warwick.ac.uk\\}
\addtocounter{footnote}{-1}\let\thefootnote\svthefootnote

\begin{abstract}
Exposing latent knowledge in geospatial trajectories has the potential to provide a better understanding of the movements of individuals and groups. Motivated by such a desire, this work presents the \emph{context tree}, a new hierarchical data structure that summarises the context behind user actions in a single model. We propose a method for context tree construction that augments geospatial trajectories with land usage data to identify such contexts.
Through evaluation of the construction method and analysis of the properties of generated context trees, we demonstrate the foundation for understanding and modelling behaviour afforded. Summarising user contexts into a single data structure gives easy access to information that would otherwise remain latent, providing the basis for better understanding and predicting the actions and behaviours of individuals and groups. Finally, we also present a method for pruning context trees, for use in applications where it is desirable to reduce the size of the tree while retaining useful information.
\end{abstract} 
 
\section{Introduction}\label{sec:introduction}

Exposing the latent knowledge present in geospatial trajectories has become an increasingly important research topic in recent years, due in part to the pervasiveness of location-aware hardware and the resulting availability of trajectory data. Motivated by a desire to understand the movement patterns of users, this paper presents a new data structure, the \emph{context tree}, that summarises the context behind user actions in a single hierarchical model. Additionally, the paper proposes a method for generating context trees from geospatial trajectories and land usage information, and provides concrete implementations for each stage of the method, namely augmentation, filtering, and clustering. A context tree itself is formed of clusters at multiple scales that describe the contexts in which the user was immersed, affording easy access to information that would have previously remained hidden, forming the basis for understanding and predicting the actions and behaviours of individuals and groups.

Existing work in understanding people through the context of activities has considered various attributes as defining context, including an individual's location, the current time and weather, and other individuals who are nearby~\cite{Dey:1999uq,Schilit:2000tf}, typically using data collected from smartphones~\cite{Bao:2011hy,Cao:2010bb,Huai:2014cm}. While existing approaches provide a basis for context-aware applications, they are limited by the data that can be collected directly from the user. Augmenting geospatial trajectories with land usage information enables the identification of contexts that consider the type and properties of the location of an activity.

In this paper we present the following contributions: (i) the \emph{context tree} data structure that hierarchically represents user contexts at multiple scales, (ii) a method for constructing context trees from geospatial trajectories and land usage information, (iii) a set of concrete techniques to achieve each stage in the construction method, namely augmentation, filtering and clustering, (iv) evaluation of context trees constructed from real-world data, and an analysis of the properties that make them amenable for use in understanding individuals, and (v) a method of pruning context trees, to reduce their size while retaining useful information. 

The remainder of this paper is structured as follows. Section~\ref{sec:related} discusses relevant related work in location extraction and activity and context identification. In Section~\ref{sec:overview} we propose the context tree, a new data structure, and present an overview of the method employed for constructing context trees. Concrete implementations of the stages of this method are given in Sections~\ref{sec:processing} and~\ref{sec:clustering}. We present an evaluation of context trees in Section~\ref{sec:evaluation}, and discuss pruning the generated trees in Section~\ref{sec:pruning}. Finally, we conclude the paper with a discussion of future work and applications in Section~\ref{sec:conclusion}.
\section{Related Work}\label{sec:related}

Geospatial trajectories, usually collected from GPS logging devices, have been used as a basis for knowledge acquisition in many areas, including for location extraction~\cite{Andrienko:2011ja,Ashbrook:2002du,Ashbrook:2003bi,Bamis:2011gn,Montoliu:2010bh,Thomason:2015ts,Thomason:2016gv}. Periods of low mobility are extracted from the trajectories and clustered using techniques such as DBSCAN~\cite{Ester:1996tm} and k-means~\cite{MacQueen:1967uv}, identifying areas in which the time was spent. These techniques identify areas of arbitrary shape, but are incapable of identifying places where non-stationary activities took place. Augmenting identified areas with additional information, Yan et al.~[\citeyear{Yan:2013ch}] propose a technique for the derivation and modelling of \emph{semantic trajectories}. However, the additional data sources are not leveraged for identifying locations, only for providing labelling after locations have been identified.

Once identified, significant locations have formed the basis for many applications, including location prediction using Markov models~\cite{Ashbrook:2002du,Ashbrook:2003bi}, neural networks~\cite{Thomason:2015vs}, periodicity-based approaches~\cite{Wang:2012ta}, and blockmodels~\cite{Fukano:2013ju}. Using multilayer perceptrons for location prediction, Thomason et al.~[\citeyear{Thomason:2015po}] evaluate extracted locations and predictions to perform automatic parameter selection for location extraction and prediction. Research has also considered predicting when a user will next visit a specific location using Bayesian inference~\cite{Gao:2012ud}, how long a user will stay at a given location~\cite{Liu:2013hp}, as well as developing techniques to apply labels in a semi-supervised manner to extracted locations to provide additional meaning~\cite{Krumm:2013fg}. Prediction has also occurred without the need for location extraction, in the form of destination prediction, achieved by identifying similar historical trajectories to a current one through clustering approaches~\cite{Chen:2010ev,Monreale:2009is,Nakahara:2012ul}, Bayesian inference~\cite{Horvitz:2006ua} and hidden Markov models~\cite{AlvarezGarcia:2010hg}. Similarly, predicting journey duration has been explored using neural networks~\cite{Chen:2009jm}, along with predicting when two people will next meet~\cite{Yu:2015je}, and providing recommendations to users new to a city based on the locations visited by others~\cite{Bao:2013ch,Zheng:2011go}.

While trajectories have also been used to identify non-stationary activities, in the form of transport mode identification through change-point detection and classification-based approaches~\cite{Liao:2007ff,Patterson:2003tc,Zheng:2008ku,Zheng:2008ii}, many related techniques operate on different sources of data. Activity detection has been achieved from video data by Kim et al.~[\citeyear{Kim:2010cb}], who use Markov models to identify the activities being performed. Unfortunately, ensuring the constant availability of video data on an individual is infeasible. Research has therefore considered identifying the activity being performed from low-level sensor data (e.g.\ accelerometers and heart-rate) generated by devices carried by individuals, using classifiers and related techniques to label periods of data from a set of possible activities~\cite{Choudhury:2008hs,Lee:2002ur,Lester:2005wt,Morris:2011bf,Pirttikangas:2006tc,Ravi:2005tw}. 

Context, situation, and intention awareness have also been considered, where a context aims to identify times when a user was performing the same task, without necessarily knowing what the task is. Literature in this domain has explored using entropy-based clustering to identify contexts~\cite{Bao:2011hy}, and sequence-based approaches that consider the transitions between contexts~\cite{Lemlouma:2004cm}. Utilising contexts, research has also focused on developing architectures and applications that adapt devices based on the current context~\cite{Anagnostopoulos:2006ge,Lemlouma:2004cm}. Situation and intention awareness is more focused on developing tools and techniques to aid a person in conducting a particular task to achieve some goal~\cite{Howard:2013ij,Vinciarelli:2015fw}, with specific examples in defence~\cite{Howard:2002tk} and aviation~\cite{Endsley:2000vk,Endsley:1995ur}. As with location extraction, however, existing techniques focus only on collected data, and do not attempt to augment this with other data available after collection. Such augmentation could offer greater insight into the entities a person was interacting with, enabling a better understanding of the actions they were performing.

Focusing on only trajectories, literature has also considered the identification of repeating patterns, both from geospatial trajectories~\cite{Cao:2005hq,Cao:2007kj,Eagle:2009jx,Giannotti:2007ha,Gudmundsson:2004ee}, and general object movement trajectories~\cite{Li:2010de,Yang:2003eu}, where repeating patterns are expected to consist of activities that the user repeatedly conducts. Such patterns have also been considered as routines, where the aim is to extract features of a given day for classification (e.g.\ ``left work at 5PM'')~\cite{Farrahi:2008ir,Farrahi:2010fz}. Patterns, and extracted location transitions, have formed the basis of user similarity identification~\cite{Xiao:2014dh}, and travel companion identification~\cite{Tang:2012wt}. Once expected patterns for a given user or group have been extracted, anomalous actions become possible to identify. Anomaly detection has been performed on geospatial trajectories, where isolation-based outlier detection has identified anomalous subtrajectories from vehicle tracking data~\cite{Chen:2012ho,Zhang:2011ba}. Similarly, statistical approaches have been shown to be useful in identifying trajectories that differ from an expected pattern~\cite{Laxhammar:2011td,Laxhammar:2014ix,Rosen:2012gh}.

Raw geospatial trajectories have been used as the basis for many different tasks and applications. While assuming the availability of additional data at time of collection is often infeasible, augmenting trajectories after collection is possible and can enrich the knowledge afforded. Applications that consider such augmented trajectories include using map data to fill in missing periods of a trajectory~\cite{Zheng:2012kz}, and using map searches augmented with trajectories from the same user to enhance destination prediction~\cite{Wu:2015ta}. While existing work by Yan et al.~[\citeyear{Yan:2013ch}] has considered the augmentation of trajectories to understand the semantics behind trajectory segments, they do not attempt to utilise the semantics to influence the partitioning of trajectories or identify contexts. Understanding the semantics behind trajectories from the beginning has the potential to better understand what a person was doing and their interactions, and thus provide a foundation for identifying similar contexts. 

\subsection{Geospatial Datasets}\label{sec:related:data}

Although it is increasingly becoming easier to collect geospatial data due to the proliferation of location-aware devices such as smartphones, the availability of public geospatial datasets still presents a challenge for researchers. Privacy concerns are the main obstacle to making such data publicly available, and consequently there are only a limited number of public datasets, each having certain drawbacks. To overcome these issues, many researchers have collected data themselves for use in their work~\cite{Ashbrook:2003bi,Thomason:2015ts,SiiaNowicka:2015dq}. However, using such private datasets decreases the reproducibility of work and thus the use of public data is preferred. Such publicly available datasets include MIT's Reality Mining dataset~\cite{Eagle:2005er}, which uses cell towers to estimate the locations of devices belonging to 100 students. More recently, GPS-enabled devices have been used for data collection to produce Microsoft's GeoLife Trajectories~\cite{Zheng:2008ku,Zheng:2009td,Zheng:2010uc}, Nokia's Mobile Data Challenge (MDC) dataset~\cite{Kiukkonen:2010vm,Laurila:2012vk}, and the Yonsei dataset~\cite{Chon:2011ki}. While all of these datasets are available for research purposes, they have their own caveats. The Yonsei dataset contains only 2 months worth of data, while the GeoLife and MDC datasets contain vast amounts of data collected over several years from hundreds of users. The GeoLife dataset is focused on when participants were moving, and therefore does not provide continuous data. The MDC dataset, on the other hand, aims to provide continuous data collected from smartphones, although for privacy reasons the areas surrounding the known residences of participants have their accuracies significantly reduced in the data. Despite this, and due to its continuous collection methodology, the MDC dataset is still one of the most accurate and representative datasets available, containing real-world data collected from the smartphones carried by nearly 200 users over a span of 2 years. The MDC dataset also includes accuracy values, indicating how much confidence can be placed in the recorded coordinates, information which is not provided with GeoLife.

\subsection{Investigative Techniques}\label{sec:related:groundtruth}

For many techniques relating to extracting knowledge from data, collecting a concrete ground truth is infeasible. Significant location extraction, for example, can extract locations at various scales and so no single ground truth can exist. Existing literature addresses this by exploring the properties of the outputs from such techniques and comparing these properties to expected results. For instance, Guidotti et al.~[\citeyear{Guidotti:2015ji}] create synthetic trajectories with known properties and devise metrics to compare extracted locations with desirable properties. Thomason et al.~[\citeyear{Thomason:2015ts}] compare properties of the identified locations against acceptable ranges of values, determined from knowledge of the input data, as well as demonstrating the applicability of the technique through examples. For travel-mode classification, Si{\l}a-Nowicka et al.~[\citeyear{SiiaNowicka:2015dq}] compare against a small set of manually labelled subtrajectories. In this paper, to cope with the limited availability of user-provided ground truth, we present an evaluation that both compares the output of the proposed approach to a limited ground truth and characterises the outputs of the algorithm through a set of metrics. While a ground truth may not exist in all domains, an understanding of the performance and applicability of the proposed approach can be achieved through characterising the outputs and manually generating or labelling subsets of the data to create a partial ground truth.
\section{Proposed Structure: The Context Tree}\label{sec:overview}

\begin{figure}[t!]
  \centering
  \includegraphics[width=0.7\linewidth]{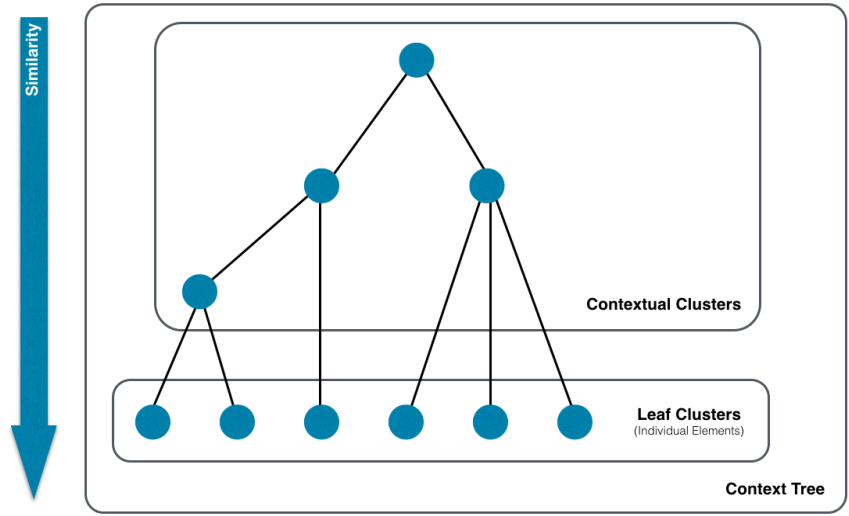}
  \caption{An abstract representation of a context tree, in which the similarity of nodes increases with depth.}\label{fig:diagram_context_tree}
\end{figure}

This paper proposes and evaluates the \emph{context tree} hierarchical data structure, that summarises the contexts that a user has been immersed within at multiple scales. Each leaf node of the tree represents a real-world feature or element that the user has likely interacted with, be it a specific building, area, or individual feature (e.g.\ a bench in a park). These individual elements are joined together through \emph{context nodes} that represent a context at a specific scale, where time spent within a context means that the user likely had similar aims or goals, and are identified by exploring time the user spends interacting with elements with similar properties, or elements that are interacted with in a similar manner. As it summarises time in this way, the context tree can become the basis for understanding people from augmented geospatial data. The context tree structure is depicted in Figure~\ref{fig:diagram_context_tree}.

\begin{figure}[p]
	\centering
	\includegraphics[width=1\linewidth]{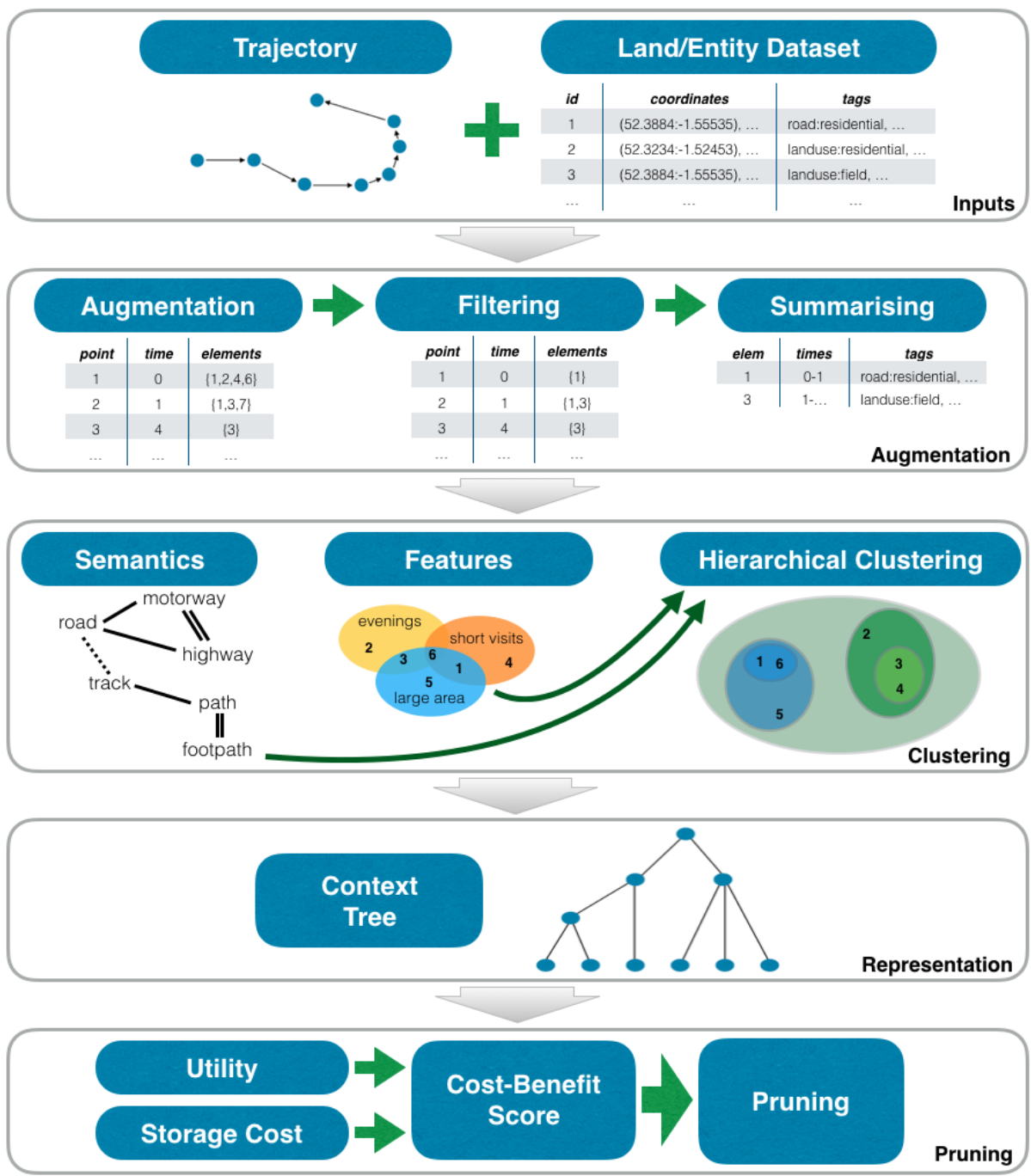}
	\caption{Overview of the context tree generation framework. A trajectory is augmented with land usage data, and this augmented data is then hierarchically clustered into a context tree. Subsequently the context tree can optionally be pruned.}\label{fig:overview}
\end{figure}

Generating a context tree requires both a geospatial trajectory and a dataset of land usage features, along with a multi-stage process for augmentation, filtering and clustering this data into a useful structure. The remainder of this paper presents the proposed method for generating context trees, provides an evaluation of context trees, and presents a method of pruning  context trees to reduce their size while maintaining information. The method for augmenting geospatial trajectories with land usage information, to summarise contexts into a context tree, consists of the following five stages, as depicted in Figure~\ref{fig:overview}.

\begin{description}
\setlength\itemsep{-1em}
\item[1. \textbf{Inputs}]\hfill \\ The raw geospatial trajectory and land usage data enters the system.\\
\item[2. \textbf{Augmentation}]
\hfill \\ Land usage elements likely to have been interacted with are identified by extracting all potential elements and filtering them to remove noise.\\
\item[3. \textbf{Clustering}]
\hfill \\ Filtered land usage elements, and their interactions, become the basis for contextual clustering. Clustering is achieved with a hierarchical agglomerative algorithm.\\
\item[4. \textbf{Representation}]
\hfill \\ Once clustered, the elements form a context tree data structure that can be used as the basis for further understanding  the behaviours of individuals and groups.\\
\item[5. \textbf{Pruning}]
\hfill \\ Some applications may be limited by the amount of data they can store, or processing they can perform, and so it may be necessary to prune a context tree to reduce its size while maintaining as much useful information as possible. Pruning is achieved through analysing the nodes of a context tree with respect to a defined set of metrics.
\end{description}

\noindent In the following sections we describe the stages of augmentation (Section~\ref{sec:processing}), clustering (Section~\ref{sec:clustering}), representation (Section~\ref{sec:clustering}), and pruning (Section~\ref{sec:pruning}) in more detail.

\section{Trajectory Augmentation}\label{sec:processing}

In order to better understand users through their past actions, and assuming only geospatial data is available at the point of data collection, this section describes the process of trajectory augmentation that combines raw trajectories with land usage data. A trajectory is a temporally ordered sequence of data points that locate an individual or entity:
$$T = (p_1, p_2, p_3, ..., p_n)$$

\noindent where $p_i = \{t_i, l_i, a_i\}$ is an individual trajectory \emph{point}, consisting of time ($t_i$), location ($l_i$, e.g.\ a $<lat,lng>$ pair) and accuracy ($a_i$, typically measured in metres).

In addition to such trajectories, land usage data can also be used for identifying locations and entities that are meaningful to the user. Land usage data is assumed to be sets of \emph{entities} with associated information. An entity, in this case, directly maps to a single real-world object, feature, or area, such as an individual postbox, field, or building. It can also refer to a collection of such entities that form a larger designation, such as a university campus or residential housing area. Each of these elements is expected to be associated with a set of geographical coordinate pairs that represent its shape and location, in addition to a set of tags in the form of `key:value' pairs that describe properties of the element, including its type and usage (e.g.\ a house may be tagged as `building:residential').

\begin{figure}[t]
  \centering
  \includegraphics[width=0.95\textwidth]{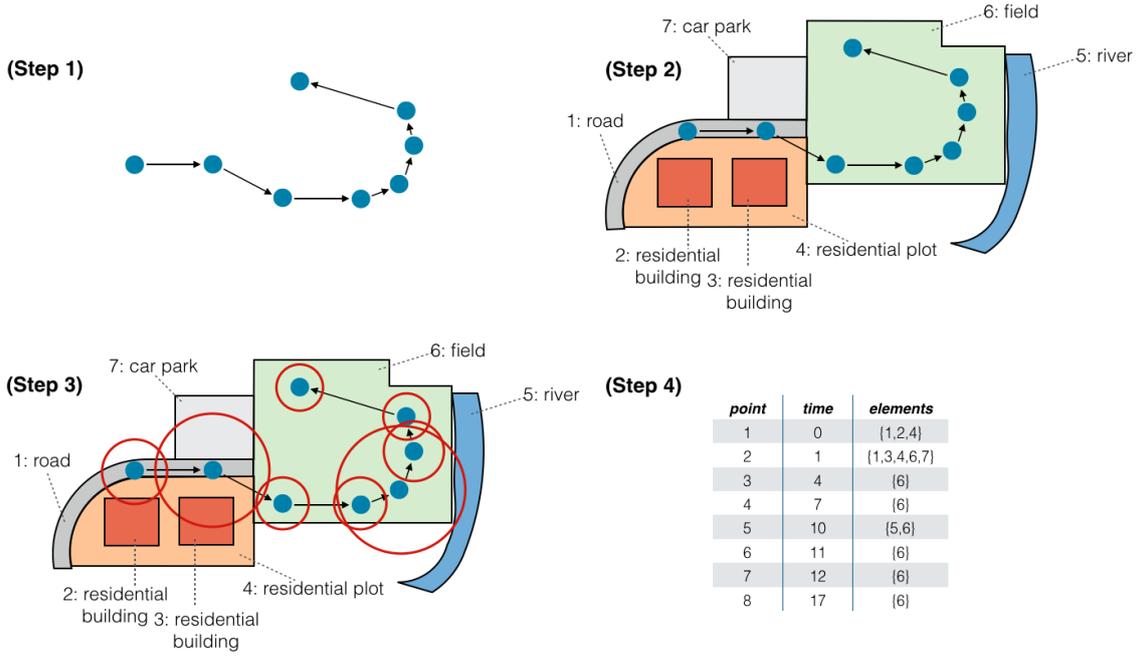} 
  \caption{Trajectory augmentation procedure.}\label{fig:diagram_trajectory_augmentation}
\end{figure}

\subsection{Element Extraction}\label{sec:element_extraction}

The process for extracting relevant land usage elements is illustrated in Figure~\ref{fig:diagram_trajectory_augmentation}. A raw geospatial trajectory (Step 1) is overlaid on a land usage dataset (Step 2), at which point the accuracy recorded by the location measuring device (e.g.\ GPS, measured in metres) is used (Step 3), such that all elements that are partially or wholly within the radius are stored alongside the original trajectory point (Step 4). This procedure is completed automatically by iterating through each trajectory point and querying the land usage dataset for any element that intersects or covers any part of the accuracy radius. 

\subsection{Filtering}\label{sec:filtering}

\begin{figure}[t]
  \centering
  \includegraphics[width=0.95\textwidth]{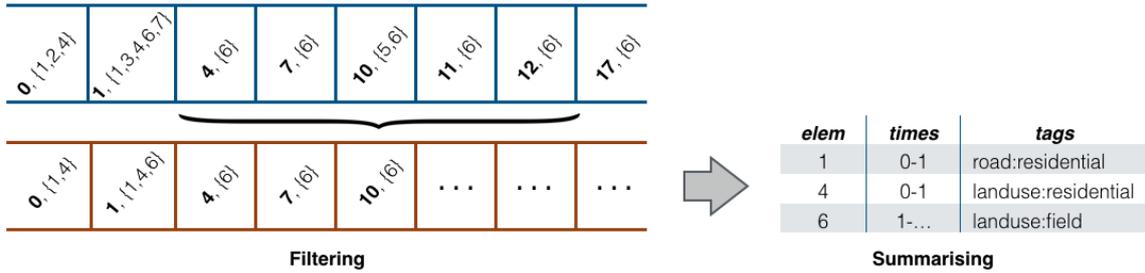} 
  \caption{Example of filtering augmented trajectories to remove noise, and subsequent summarising through clustering of contiguous time periods.}\label{fig:diagram_filtering}
\end{figure}

\noindent As the element extraction process (Section~\ref{sec:element_extraction}) augments trajectories with all land usage elements that fall within the \emph{accuracy radius} of a trajectory point, it is prone to including a significant number of elements with which the user was not interacting. To cope with these noise elements a filtering procedure can be used. Our proposed filter is a generalised version of a weighted average filter, a technique typically used to smooth noisy signals, modified to operate over sets of land usage elements and depicted in Figure~\ref{fig:diagram_filtering}. The filter maintains a buffer of elements and selects from this buffer based on an assigned weight in a three-step process:

\begin{enumerate}
\item A buffer of points, and associated land usage data, is selected.
\item The land usage elements in the buffer are weighted and scored.
\item Elements are selected, based on their score, for inclusion in the output.
\end{enumerate}

\subsubsection{Buffer Selection}

Due to the nature of geospatial data collection systems, a continuous and evenly timesliced trajectory cannot be assumed, and so selecting a buffer based on a fixed number of points would be inappropriate. Instead, we use a fixed temporal width for the buffer and consider all points that fall within this period. A buffer therefore consists of a \emph{point under consideration}, and the points falling within $\delta$ seconds immediately before or after this point. The pseudocode for maintaining such a buffer is presented in Algorithm~\ref{alg:buffer_management}.

\begin{algorithm}[t!] 
\caption{Buffer Management}\label{alg:buffer_management}
\algrenewcommand\alglinenumber[1]{\scriptsize #1:}
\algrenewcomment[1]{\(//\) #1}
\begin{algorithmic}[1]
\small

\State \textit{points} $\gets (p_1, p_2,...)$ \Comment input set
\State $\delta \gets$ 300 \Comment input parameter specifying buffer width
\State \textit{buffer} $\gets$ [ \textit{points}.shift ]
\State \textit{output} $\gets$ [~]
\State \textit{index} $\gets$ null\\

\\\Comment Build the initial buffer
\While{\textit{points}.length $> 0$}\\
~~~~\Comment If \textit{index} has not been set, then we are in the first half
    \If{\textit{index} $==$ null \&\& TimeBetween(\textit{buffer}[0], \textit{points}[0]) $> \delta$}\\
~~~~~~~~\Comment If the next point is greater than $\delta$ seconds from the first, then the first half is full
        \State \textit{index} $\gets$ \textit{buffer}.length $-$ 1\\
~~~~\Comment If \textit{index} has been set, then we are in the second half
    \ElsIf{\textit{index} $!=$ null \&\& TimeBetween(\textit{buffer}[index], \textit{points}[0]) $> \delta$}
        \State break \Comment Exit the loop as adding the next point would exceed $\delta$
    \Else
        \State \textit{buffer}.append(\textit{points}.shift)
    \EndIf
\EndWhile\\

\\\Comment Process the current buffer, increment $index$ and maintain the new buffer
\While{\textit{points}.length $> 0$}
    \State \textit{output}.append(Filter(\textit{buffer}, \textit{index})) \Comment Perform the actual filtering
    \State \textit{index} $\gets$ \textit{index} + 1\\

\\~~~~\Comment If the point for consideration is not in the buffer, then add it now
    \If{\textit{index} $==$ \textit{buffer}.length}
        \State \textit{buffer}.append(\textit{points}.shift)
    \EndIf\\

\\~~~~\Comment Remove any point from the first part that is not within $\delta$ seconds of \textit{buffer}[index]
    \While{TimeBetween(\textit{buffer}[0], \textit{buffer}[\textit{index}]) $> \delta$}
        \State \textit{buffer}.shift
        \State \textit{index} $\gets$ \textit{index} - 1
    \EndWhile\\

\\~~~~\Comment Add points until doing so would exceed $\delta$ seconds from buffer[index]
    \While{\textit{points}.length $>$ 0 \&\& TimeBetween(\textit{buffer}[index], \textit{points}[0]) $<= \delta$}
        \State \textit{buffer}.append(\textit{points}.shift)
    \EndWhile

\EndWhile \\

\State \Return \textit{output}

\end{algorithmic}
\end{algorithm}

\subsubsection{Scoring}

Scores are then applied to each land usage element in the buffer, weighted by the number of points the element is associated with, the accuracy of these points and the temporal distance from the point under consideration. Since we are dealing with sets, rather than the filter simply averaging values over the buffer, the process is modified by assigning weighted scores to each set element and then selecting elements according to a threshold. Combining these factors into a score, we have:
\begin{equation}\label{eqn:score} 
Score(e) = \sum_{p\in P_e}\left(\frac{1}{a_p} \times \left(1-\frac{dist(p, p_c)}{\delta}\right)\right) \times |P_e| 
\end{equation}

\noindent where $P_e$ is the set of all points that are associated with element $e$, $a_p$ is the accuracy value of point $p$ (in metres, such that low values indicate that there is likely to be less noise), $p_c$ is the point under consideration, $\delta$ is the width of the buffer (i.e.\ the maximum number of seconds from $p_c$ to consider) and $dist(p_1,p_2)$ is the number of seconds between points $p_1$ and $p_2$ (temporal distance). Equation~\ref{eqn:score} gives a higher score to elements associated with a large number of high accuracy points (where high accuracy is recorded as a small value). Scores are then normalised relative to the maximum:
\begin{equation}\label{eqn:normScore} 
NormalisedScore(e) = \frac{Score(e)}{argmax_{Score}(Score(e) : \forall e \in \mathit{buffer})}
\end{equation} 

\subsubsection{Selection}\label{sec:filtering:selection}

With each element in the buffer assigned a score, selection can occur either by using a fixed threshold to discard low-scoring elements, or by keeping all elements but limiting their effect through soft-thresholding. Soft-thresholding is a technique commonly applied in signal processing, where a kernel is applied to the calculated scores, forcing higher scores closer to 1 and lower scores closer to 0. While soft-thresholding removes the need to apply a fixed threshold, for this work we are only concerned with whether or not an element is included in the output set and thus we employ a threshold, $t$, where any element with a $NormalisedScore$ of greater than $t$ becomes part of the output set, and the remaining elements are discarded.

\subsection{Data Summarisation}\label{sec:summarisation}

Once filtered, augmented trajectories contain a record of where an individual was at a given time, along with the real-world features they were likely interacting with. These interactions are summarised into continuous spans of time by considering each land usage element encountered. If the same land usage entity is associated with two consecutive points, it can be assumed that it is also associated with the period of time between these points, if such a period of time is sufficiently small. An example summary is shown in Figure~\ref{fig:diagram_filtering} (right).

If the time between consecutive points is large, it cannot be known whether the user ceased interacting with an element and resumed again before data collection next occurred, and so a limit on the time between consecutive points is specified as $t_{max}$. If the time between two consecutive points associated with the same element is greater than $t_{max}$, then the interactions are split, a technique also used in location extraction applications~\cite{Montoliu:2010bh,Thomason:2016gv}. This results in a summary list of land usage elements along with a set of times, during which the individual can be assumed to have been interacting with the element in question.
\section{Contextual Clustering}\label{sec:clustering}

The identification of similar \emph{contexts} is performed through clustering that considers both the properties of the elements and the properties of user interactions to determine similarity. Rather than aiming to identify a single level of clusters, which would limit the utility and applicability of the clusters to a single scale, the goal here is to build a hierarchical model, constructed by progressively merging land usage elements that represent similar contexts in a context tree, a depiction of which is shown earlier in Figure~\ref{fig:diagram_context_tree}. 

\subsection{Building Clusters}\label{sec:clustering:merging}

Initially, each land usage element is distinct and is treated as a singleton cluster (i.e.\ a cluster with exactly one element). At each round of clustering, several of these clusters are merged to represent a context and a new higher level in the hierarchy, with pointers between the levels considered as \emph{parent} and \emph{child} relationships. That is, if two clusters at one level become merged into another cluster at the next level, the original clusters are considered as \emph{children} of the new cluster. This section describes how clusters are merged with respect to their properties.

As discussed in Section~\ref{sec:processing}, land usage elements are assumed to have a set of \emph{tags} in the form of `key:value' pairs that describe properties of the real-world entity to which the element relates, in addition to \emph{geographical coordinate sets} that describe the geographical properties of the real-world entity. Once augmented and summarised, these elements are also associated with a set of \emph{times}. When clusters are merged to create a context tree, the following procedures are used:

\begin{description}
\item[\textbf{Times}]\hfill \\The times for the merged cluster are taken to be the union of the sets of times from all child clusters, where overlapping time ranges are themselves combined into one. For example, if one cluster had the set of times \{10:00-10:05, 11:00-12:00\} and another had \{10:04-10:20, 11:10-11:15, 12:05-12:09\}, then the merged times would be \{10:00-10:20, 11:00-12:00, 12:05-12:09\}.\\
\item[\textbf{Tags}]\hfill \\ Similarly, each element has associated tags. The tags of the merged cluster are defined as the union of tags from the child clusters, where if two tags share a key but not a value, both values are stored.\\
\item[\textbf{Geographical Coordinate Sets}]\hfill \\ Each element contains a set of coordinates that define the geographical shape of the entity to which they relate. Merging such elements should keep each of these sets discrete, unless they intersect in which case the coordinates belonging to both shapes are combined and replaced with their convex hull.
\end{description}

\begin{figure}[t]
  \centering
  \includegraphics[width=0.9\linewidth]{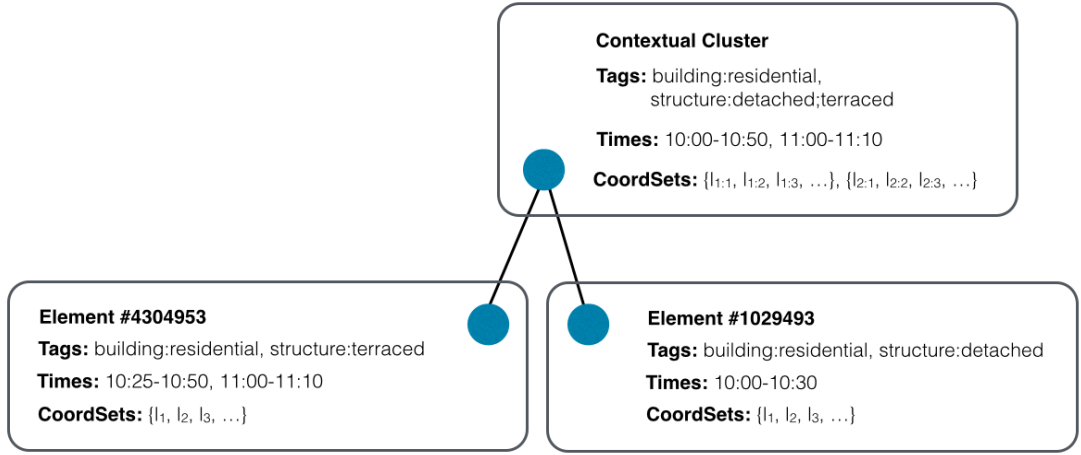}
  \caption{Cluster merging example.}\label{fig:diagram_merging}
\end{figure}

\noindent The merging of \emph{Times} assumes a periodicity of 24 hours, which while reasonable for many people (i.e. those who follow a daily routine), it may not be appropriate for everyone. As such, automatic time series learning could be utilised to better learn meaningful movement patterns of the individual. While exploring such techniques is beyond the scope of this paper, there are many existing approaches that may be effective for the task~\cite{Ahmad:2004bg}. An example merging of two elements according to these rules is shown in Figure~\ref{fig:diagram_merging}, where it is assumed that there is no geographical overlap between the two elements (i.e.\ the coordinate sets cannot be merged).

\subsection{Contextual Distance Metrics}\label{sec:clustering:similarity} 

Clustering elements together requires a distance metric to measure element similarity. While identifying contexts from certain types of data is a task considered before, and discussed in Section~\ref{sec:related}, no metrics currently exist that have been tailored to the identification of contexts from augmented geospatial trajectories. This section presents metrics that encapsulate the goals behind context extraction for this specific problem, with an emphasis on properties of the interactions and properties of the real-world features being interacted with. Having defined how elements are merged into clusters and, consequently, how two clusters are merged (Section~\ref{sec:clustering:merging}), we can now consider the similarity between two clusters.

\subsubsection{Semantic Similarity}\label{sec:cluster:semantic_similarity}

Clusters have tags that describe properties of the real-world entities contained in the cluster, forming an ideal basis for understanding what the user might have been doing. Under the assumption that clusters with similar tags are likely to have properties in common, we use the semantic similarity between cluster tags as the basis for a distance metric. For this, we adopt the similarity measure proposed by Wu and Palmer~[\citeyear{Wu:1994do}], and extended by Resnik~[\citeyear{Resnik:1999ub}] for calculating distance between word taxonomies through WordNet~\cite{Miller:1995co}. The calculated scores are between 0 and 1 (inclusive), where a score of 1 means that the words are interchangeable. The semantic similarity between two sets of tags, $t_1$ and $t_2$, is therefore calculated as:
\begin{equation} 
TagSim(t_1, t_2) = \frac{\sum_{t \in t_1} {argmax_{Sim}(Sim(t, t_{21}), Sim(t, t_{22}), ..., Sim(t, t_{2i}))}}{|t_1|}
\end{equation} 

\noindent As tag similarity is not commutative, cluster similarity is calculated as:
\begin{multline}
SemanticSimilarity(c_1, c_2) = \\
argmax_{TagSim}(TagSim(c_1.tags, c_2.tags), TagSim(c_2.tags, c_1.tags))
\end{multline}

\subsubsection{Feature Similarity}

The context of an activity or period of time is dependent not only on the location in which time is spent, but on additional factors. With this in mind, we propose a second similarity measure, \textit{FeatureSimilarity}, that compares the interaction features of two clusters, specifically:

\begin{itemize}
\item Average interaction duration
\item Most common time of day interaction begins
\item Count of the number of times the element is interacted with
\item Total area covered by elements (in $m^2$)
\end{itemize} 

\noindent The value from each feature is then discretised by placing values within bins (e.g.\ time of day could be recorded in 4 hour increments), and converted into a single string that describes the feature and value (e.g.\ `timeofday\_12' would indicate that the most common time of day that interaction begins is between 12PM--4PM). This procedure generates a set of features, $f_1$ and $f_2$, for clusters $c_1$ and $c_2$, from which a similarity score is defined using the Jaccard index~\cite{Rajaraman:2011aa}:
\begin{equation}
FeatureSimilarity(f_1,f_2) = {\frac{|f_1 \cap f_2|}{|f_1 \cup f_2|}}
\end{equation}

\subsubsection{Geographical Distance}

For some applications it is possible that the similarity between clusters depends upon their geographical proximity, where two clusters that are close together may have common purposes. If this property is known to be true in the data, or given the goal of clustering, then the proximity of clusters can be considered as the minimum geographical distance between elements of a cluster, calculated using the Haversine formula~\cite{Robusto:1957td}:
\begin{equation}
GeographicalDistance(c_1, c_2) = argmin_{distance}(distance(x_1 \in c_1, y_1 \in c_2),\ldots)
\end{equation}

\subsubsection{Hybrid Contextual Distance}

Using one of the previously discussed metrics in isolation would not accurately capture the context of the individual, as context depends on more than just any one factor. Instead, we combine the \textit{SemanticSimilarity} and \textit{FeatureSimilarity} scores into \emph{Hybrid Contextual Distance (HCD)}, a measure of the contextual similarity between two clusters:
\begin{multline}\label{eqn:hcd} 
HCD(c_1, c_2) = \\
1 - (\lambda \times SemanticSimilarity(c_1, c_2) + (1 - \lambda) \times FeatureSimilarity(c_1, c_2))
\end{multline}

\noindent where $\lambda$ is a user-specified weighting parameter that allows emphasis to be placed either on the semantic or feature similarity between clusters. We chose to ignore the geographical proximity of elements because contexts should be separate from their geographical location (e.g. visiting two cafes in different cities is likely to be indicative of the same context). If, however, additional domain knowledge is available that ties geographical locations together with additional meaning (e.g. it is known that all buildings in a given area perform a similar function), then geographical distance could be added to the HCD metric.  HCD can  be used as a basis for clustering elements, and thus determining which elements have similar contexts, aiding in our understanding of the individual to which the data belongs.

\subsection{Hierarchical Clustering}\label{sec:clustering:algorithm}

\begin{algorithm}[t]
\caption{Agglomerative Hierarchical Clustering Algorithm}\label{alg:clustering}
\algrenewcommand\alglinenumber[1]{\scriptsize #1:}
\algrenewcomment[1]{\(//\) #1}
\begin{algorithmic}[1]
\small

 \State \textit{clusters} $\gets$ \textit{elements} \Comment The input set of \textit{elements}, each treated as its own cluster

 \While{\textit{clusters}.length $>$ 1}\\
 	\\
 	~~~~\Comment Create an $n\times n$ matrix of distances between clusters
 	\State \textit{distanceMatrix} $\gets [~[d_{11}, ...], [d_{21}, ...], ...]$ \\
 	\\
 	~~~~\Comment Find all pairs of clusters with the smallest distance between them\\
 	~~~~\Comment If multiple pairs overlap (i.e.\ share a cluster), then group them together
 	\State \textit{closestGroups} $\gets$ ClosestGroups(\textit{distanceMatrix})\\
 	\\
 	~~~~\Comment Merge each extracted group into a single cluster
 	\For{\textit{group} $\in$ \textit{closestGroups}}
 		\State newCluster $\gets$ Merge(\textit{group})\\
 		\\
 		~~~~~~~~\Comment Set the old clusters as children of the new and remove the old clusters from \textit{clusters}
 		\For{\textit{cluster} $\in$ \textit{group}}
 			\State \textit{newCluster}.children.append(\textit{cluster})
 			\State \textit{clusters}.delete(\textit{cluster})
 		\EndFor\\
 		\\
 		~~~~~~~~\Comment Add the merged cluster to \emph{clusters}
 		\State \textit{clusters}.append(\textit{newCluster})
 	\EndFor
 	\\
 \EndWhile\\
 \\
 \Comment By this point, \textit{clusters} contains a single root cluster for the hierarchy\\
 \Return \textit{clusters}.first
\end{algorithmic}
\end{algorithm}

With a distance metric in place, clustering can be performed using standard techniques. While traditional clustering is limited in that it extracts clusters at a single scale, which may not be appropriate for a given task, hierarchical clustering identifies clusters at multiple scales. We use a greedy hierarchical agglomerative clustering algorithm, presented in Algorithm~\ref{alg:clustering}, that extracts clusters of increasing similarity up to a single root node, creating a \emph{context tree}. While the hierarchical agglomerative clustering algorithm is fairly standard in itself, its application to the generation of context trees is novel. The algorithm deviates slightly from existing hierarchical clustering approaches in that it is capable of extracting multiple clusters together in a single step if they have the same distance.
\section{Evaluation and Results}\label{sec:evaluation}

In this work it is not practical to obtain a concrete ground truth to act as a point of comparison for evaluation because the \emph{correctness} of an extracted set of clusters depends on the task for which the clusters will be used. In light of this, we opt to evaluate the proposed techniques with an approach similar to those followed in existing literature where a single ground truth does not exist, as discussed in Section~\ref{sec:related:groundtruth}. This is achieved by exploring the properties of the generated context trees and comparing them against expected results while providing small, representative, examples that demonstrate the utility afforded by these procedures. 

This section evaluates the proposed context tree data structure, along with the generation method proposed in Sections~\ref{sec:processing} and \ref{sec:clustering}. Although there are many use-cases for context trees, including as a basis for anomaly detection, location prediction and city planning, we focus on understanding the high-level behaviours of an individual throughout a 24 hour period as a representative example.

\begin{figure}[t] 
  \centering
  \begin{lstlisting}[language=yaml,basicstyle=\small]
  (Step 1)
      latlng: 52.3834499, -1.56026223
      timestamp: 2013-11-08 14:09:51.000000000 Z
      accuracy: 65.0

  (Step 2)
      latlng: 52.3834499, -1.56026223
      timestamp: 2013-11-08 14:09:51.000000000 Z
      accuracy: 65.0
      data: [n_312873295, n_552101208, n_695942926, n_1014585845, n_1014585853, 
      ;  w_92341980, w_92342116, w_145179860, w_145179863, w_145179883, 
      ;  w_273005393, w_303748830, w_329376738, w_329376739, r_2437023, ...]

  (Step 3)
      latlng: 52.3834499, -1.56026223
      timestamp: 2013-11-08 14:09:51.000000000 Z
      accuracy: 65.0
      data: [w_145179860, r_2437023]

  (Step 4)
      w_145179860:
        tags:
          building: university
          building_levels: 3
        members: [n_1586185863, n_1586185883, n_727382425, n_1586185856, ...]
        times:
          - begin: 2013-11-08 13:13:05.000000000 
            end: 2013-11-08 17:16:47.000000000 
        latlngs:
          - 52.3837765, -1.5601465
          - 52.3838285, -1.5600527
          - ...
      r_2437023:
        tags: ...
  \end{lstlisting}

  \caption{Examples of the data at each stage of the augmentation and filtering processes.}\label{fig:eval:data}
\end{figure}

Figure~\ref{fig:eval:data} shows sample data at each stage of the augmentation and filtering process. Raw trajectory data, in the form of an ordered array of points (Step 1) enters the system. Each point has timestamp, longitude, latitude and accuracy values. Step 2 augments the trajectory with identifiers for all land usage elements that the user could have been interacting with at that time (as described in Section~\ref{sec:processing}). This is achieved by extracting all land usage elements within the radius of the accuracy of the point and storing the identifier of each element. Step 3 shows the augmented trajectory once filtered (as described in Section~\ref{sec:filtering}), which reduces the number of elements associated with each point, with the goal of limiting them to the elements likely being interacted with. Finally, summarisation occurs, clustering together contiguous time periods that belong to the same element (as described in Section~\ref{sec:summarisation}), shown in Step 4.

\begin{figure}[t]
   \centering
   \includegraphics[width=0.7\linewidth]{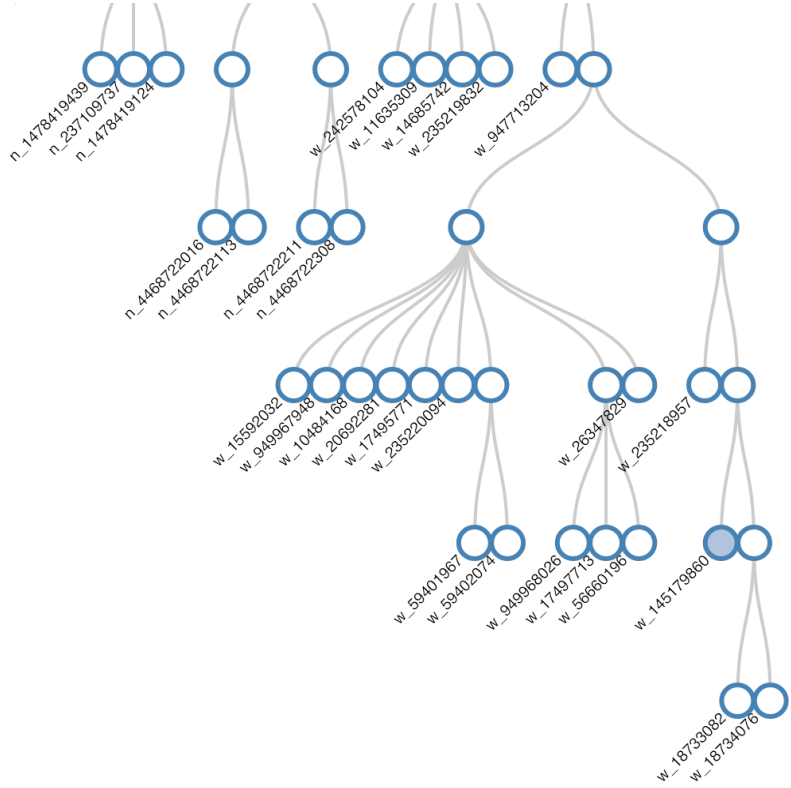}
   \caption{An extract of a tree generated using real data. The element used in the previous data examples is located in the bottom right of the image, clustered with other university buildings.}\label{fig:eval:partial_tree}
\end{figure}

Once a summarised dataset has been created, a context tree can be generated using the metrics and algorithm presented in Section~\ref{sec:clustering}. Generating a context tree from 24 hours of data produces a fairly large tree, an extract of which is shown in Figure~\ref{fig:eval:partial_tree}. A more in-depth analysis of the clustering procedure is presented in Section~\ref{sec:eval:trees}.

\subsection{Data}\label{sec:data}

Evaluating this work requires both geospatial trajectories and land usage information. The trajectories used are taken from the Nokia Mobile Data Challenge (MDC) Dataset~\cite{Kiukkonen:2010vm,Laurila:2012vk}, as discussed in Section~\ref{sec:related:data}. From this dataset, we select the real-world data from 40 users with the largest number of trajectory points for this evaluation. While this dataset contains a vast amount of information, we only consider the timestamp, latitude, longitude and accuracy of each data point (consistent with the discussion in Section~\ref{sec:processing}). In addition to this, and for comparative purposes, we also select 5 users from the GeoLife dataset~\cite{Zheng:2008ku,Zheng:2009td,Zheng:2010uc} for evaluation. While the MDC dataset aims to provide continuous coordinates for the users, the GeoLife data instead only captures periods of times when the users were moving. It has the additional drawback of not including accuracy values, which are required for this work. As the data was collected using GPS-enabled devices, we opt to assume a constant accuracy of 10m for each coordinate, in line with the expected performance of GPS~\cite{Cao:2009fh}. The trends presented in this section are consistent across both the MDC and GeoLife data, and so most of the GeoLife results are omitted for brevity, however an example can be found in Section~\ref{sec:eval:trees}.

A major drawback of using research datasets is that licences often prevent the publication of details that can be used to identify people or specific locations visited. Additionally, it is not possible to contact the users about whom data was collected to perform a user study. To get around these issues, we also collect a small dataset of our own. Aiming to match the methodology of the MDC data, trajectories were collected from the smartphones of 3 members of the Department of Computer Science, University of Warwick for a period of 3 days. These trajectories are used for illustration, instead of the MDC data, in Section~\ref{sec:eval:gt} where the presented results contain the names of specific locations visited, and communication with the users was required. 

Land usage information comes from OpenStreetMap (OSM)\footnote{\url{https://openstreetmap.org/}}, a community-maintained map that contains information pertaining to real-world entities, including their geographical coordinates and a set of tags that describe the entity. These entities include features such as individual items (e.g.\ a payphone or postbox), through to buildings and general land-usage designations (e.g.\ `farmland'). The data is extremely detailed and accurate, spanning the entire world in a consistent manner, and thus forms an ideal basis for this work. The required elements are extracted from OSM through the Overpass API\footnote{\url{https://wiki.openstreetmap.org/wiki/Overpass_API} --- By default, the Overpass API is only capable of extracting elements that a coordinate pair is contained within if the element has been assigned a name. The API has therefore been modified to consider all enclosing elements in these cases.}

\subsection{Filtering}\label{sec:eval:filtering}

The first stage in context tree generation is augmenting and filtering land usage elements. This section evaluates and characterises the performance of the filter on real-world data, by first exploring how element weights are distributed and then showing how this impacts the land usage elements that are filtered. 

\begin{figure}[t]
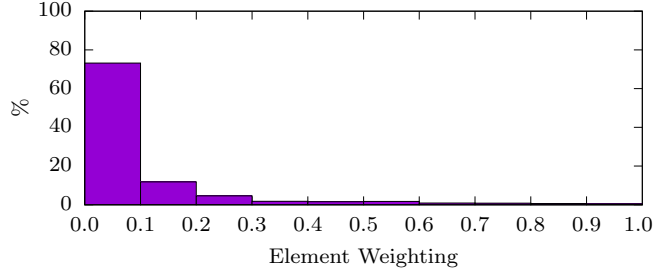

  \centering
  \footnotesize
  \begin{gnuplot}[terminal=epslatex, terminaloptions={size 3.5,1.5 font 8}]
    set xlabel "Element Weighting"
    set ylabel "$\\
    set key off
    set yrange [0:100]
    set style data histogram
    set style histogram cluster gap 0.1
    set style fill solid border -1
    set label 1 "1.0" center offset 43,-1
    set offset -1,0,0,0
    plot "data/filtering-6073-thresholding-raw.dat" using 2:xtic(1)  # with boxes #lt rgb "blue" lw 2
  \end{gnuplot}
  \caption{Distribution of element weights before filtering for an example user ($\delta = 1200$).}\label{fig:eval:thresholding:raw}
\end{figure}

\begin{figure}[t]
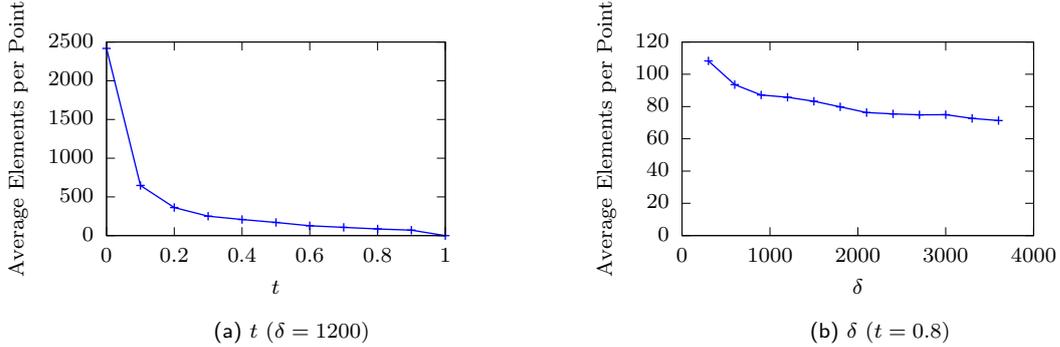

  \centering
  \footnotesize
  \begin{subfigure}[b]{0.48\textwidth}
  \begin{gnuplot}[terminal=epslatex, terminaloptions={size 2.5,1.5 font 8}]
    set key off
    set ylabel "Average Elements per Point"
    set xlabel "$t$"
    set yrange [0:]
    plot "data/filtering-6073-t.dat" using 1:2 with linespoints lt rgb "blue" lw 2
  \end{gnuplot}
  \caption{$t$ ($\delta = 1200$)}\label{fig:eval:filter:t}
  \end{subfigure}
  \begin{subfigure}[b]{0.48\textwidth}
  \begin{gnuplot}[terminal=epslatex, terminaloptions={size 2.5,1.5 font 8}]
    set key off
    set ylabel "Average Elements per Point"
    set xlabel "$\\delta$"
    set yrange [0:]
    set xrange [0:4000]
    set xtics(0,1000,2000,3000,4000)
    plot "data/filtering-6073-delta.dat" using 1:2 with linespoints lt rgb "blue" lw 2
  \end{gnuplot}
  \caption{$\delta$ ($t = 0.8$)}\label{fig:eval:filter:delta}
  \end{subfigure}
  \caption{Effect of parameters on average number of elements per point post-filtering for an example user.}
\end{figure}

Filtering takes two parameters: $\delta$ and $t$. The parameter $\delta$ specifies the width of the buffer, in seconds, and $t$ specifies a threshold where elements with a calculated weight of greater than $t$ form the output set. Holding $\delta = 1200$, Figure~\ref{fig:eval:thresholding:raw} shows the distribution of weights for all elements in the filtering process (i.e. the values of \textit{NormalisedScore} from Equation~\ref{eqn:normScore}) for all 11,575 trajectory points belonging to a sample MDC user. The effects of $t$ (with $\delta = 1200$) and $\delta$ (with $t = 0.8$) on the average number of elements per point post-filtering can be seen in Figures~\ref{fig:eval:filter:t} and~\ref{fig:eval:filter:delta}, respectively. These results are consistent with expectations, as increasing $t$ sets a higher threshold for elements to be included in the output set, and thus results in fewer elements. Increasing $\delta$ results in a greater time span considered by the filtering process, and so more elements are considered as transient, and are thus removed. Each of these figures is generated from 7 months of data from a single sample user, and while the exact numbers  vary when using data from different users, the trends remain consistent across users from both datasets.

\begin{figure}[t]
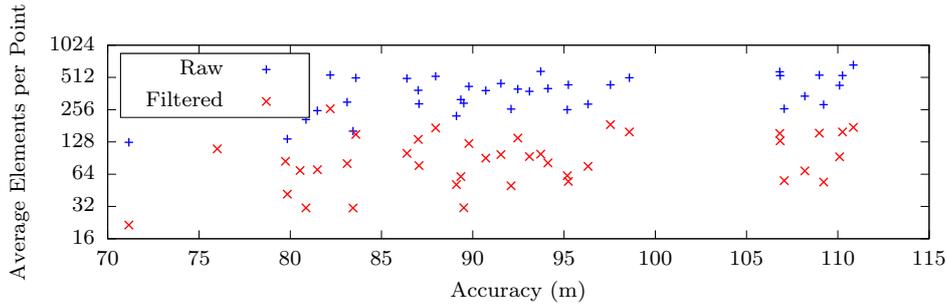

  \centering
  \footnotesize
  \begin{gnuplot}[terminal=epslatex, terminaloptions={size 5,1.5 font 8}]
    set logscale y 2
    set key left box opaque
    set key spacing 1.5
    set ylabel "Average Elements per Point"
    set xlabel "Accuracy (m)"
    plot "data/filtering-accuracy-unsorted.dat" using 1:($2/4) with points lt rgb "blue" lw 2 t 'Raw', "data/filtering-accuracy-unsorted.dat" using 1:($3/4) with points lt rgb "red" lw 2 t 'Filtered'
  \end{gnuplot}
  \caption{Effect of accuracy on number of elements, pre- and post-filtering, for different users' data.}\label{fig:eval:filter:accuracy}
\end{figure}

The accuracy of the trajectory points determines the radius of land usage data to consider. The effect of accuracy on the number of extracted elements, both pre- and post-filtering, is shown in Figure~\ref{fig:eval:filter:accuracy} for each of the 40 MDC users. The figure demonstrates that a larger accuracy typically results in a larger number of elements per point, and that filtering reduces this number.

\subsubsection{Filtering Characterisation}

\begin{figure}[t]
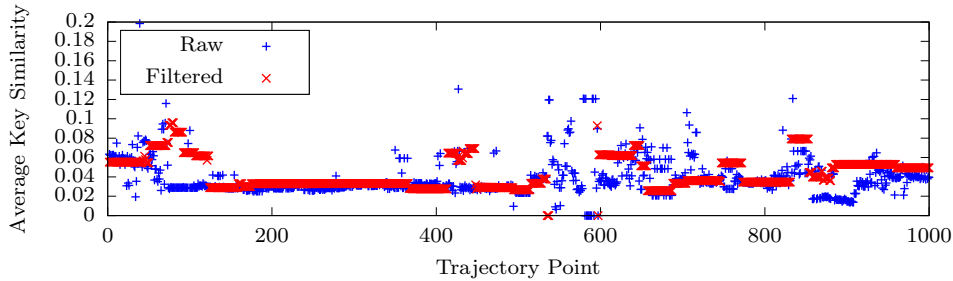

  \centering
  \footnotesize
  \begin{gnuplot}[terminal=epslatex, terminaloptions={size 5,1.5 font 8}]
    set yrange [0:]
    set xrange [0:1000]
    set ylabel "Average Key Similarity"
    set xlabel "Trajectory Point"
    set key left box opaque
    set key spacing 1.5
    plot "data/filtering-6041-convergence.dat" u 1:($4) with points lt rgb "blue" lw 2 t 'Raw', "data/filtering-6041-convergence.dat" u 1:($5) with points lt rgb "red" lw 2 t 'Filtered'#, f(x) with line lt rgb "blue" lw 2 notitle, g(x) with line lt rgb "red" lw 2 notitle
  \end{gnuplot}
  \caption{Effect of filtering on tag key similarity, both pre- and post-filtering.}\label{fig:eval:filter:convergence}
\end{figure}

To better understand the filtering process, we explore properties of the filtered data, specifically focusing on how the elements and their semantics change. The aim of filtering is to remove noise and focus the data on elements that the user was likely interacting with at a given time. It is reasonable therefore to assume that the elements post-filtering should have more similarity than those before, with less variation caused by the inclusion of \emph{random} elements. To explore this hypothesis, Figure~\ref{fig:eval:filter:convergence} shows the average tag key similarity (i.e.\ only the key part of the `key:value' pair that makes up an element's tags, which corresponds to broad type, e.g.\ `building') both pre- and post-filtering for a given user over 1000 points of their data. This  demonstrates that in the majority of cases, tag key similarity is increased, and variance significantly reduced, after filtering has occurred, indicating that the elements present post-filtering are more similar and that unrelated noise elements have been correctly removed. The semantic similarity of these tags is calculated using the method proposed in Section~\ref{sec:cluster:semantic_similarity}.

\subsection{Summarising Data}\label{sec:eval:summarising}

\begin{figure}[t]
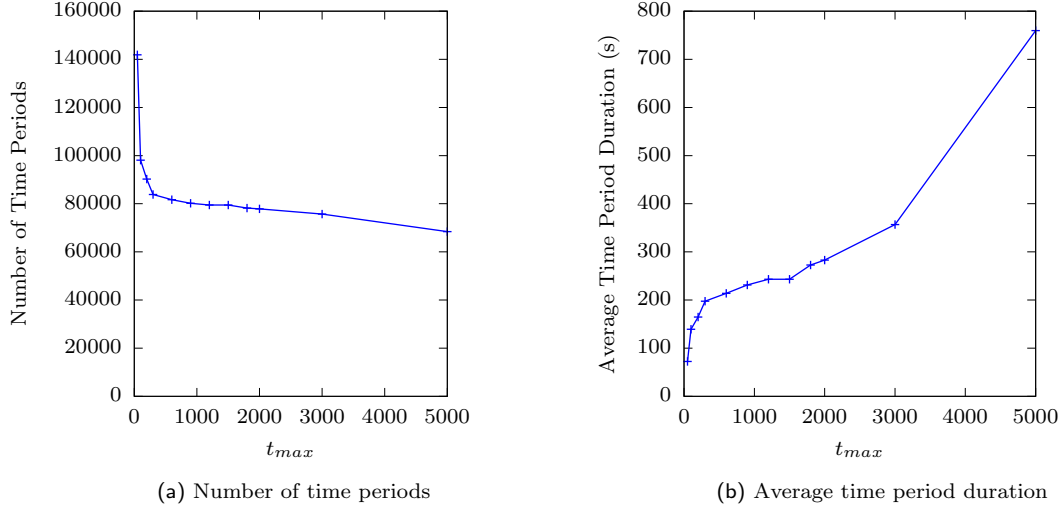

  \centering
  \footnotesize
  \begin{subfigure}[b]{0.48\textwidth}
  \begin{gnuplot}[terminal=epslatex, terminaloptions={size 2.5,2.5 font 8}]
    set key off
    set ylabel "Number of Time Periods"
    set xlabel "$t_{max}$"
    set yrange [0:]
    set xrange [0:]
    set xtics(0,1000,2000,3000,4000,5000)
    plot "data/summarising-6073-tmax.dat" u 1:2 with linespoints lt rgb "blue" lw 2
  \end{gnuplot}
  \caption{Number of time periods}\label{fig:summarising:tmax:count}
  \end{subfigure}
  \begin{subfigure}[b]{0.48\textwidth}
  \begin{gnuplot}[terminal=epslatex, terminaloptions={size 2.5,2.5 font 8}]
    set key off
    set ylabel "Average Time Period Duration (s)"
    set xlabel "$t_{max}$"
    set yrange [0:]
    set xrange [0:]
    set xtics(0,1000,2000,3000,4000,5000)
    plot "data/summarising-6073-tmax.dat" u 1:3 with linespoints lt rgb "blue" lw 2
  \end{gnuplot}
  \caption{Average time period duration}\label{fig:summarising:tmax:duration}
  \end{subfigure}
  \caption{Effect of $t_{max}$ on the summarising procedure.}
\end{figure}

Once the data has been filtered, it is summarised into continuous periods of time. Only one parameter, $t_{max}$, is required, specifying the maximum amount of time (in seconds) between consecutive points for them to be considered contiguous. Using the parameters $\delta = 1200$ and $t = 0.8$, Figure~\ref{fig:summarising:tmax:count} shows how $t_{max}$ affects the number of such periods extracted, and Figure~\ref{fig:summarising:tmax:duration} shows how $t_{max}$ affects the average length of such periods.

Figures~\ref{fig:summarising:tmax:count} and~\ref{fig:summarising:tmax:duration} show that increasing $t_{max}$ causes fewer, but longer, time periods to be extracted. In the remainder of this paper, we use $t_{max} = 1200$ (i.e.\ 20 minutes) as it provides enough time for a user to have ceased interacting with an element and to have later recommenced interaction, without causing too many interactions to be needlessly split. 

\FloatBarrier
\subsection{User-informed Evaluation}\label{sec:eval:gt}

While there is no ground truth available for this type of problem, we can evaluate the procedure by considering desirable properties of the output for a specific application and manually compare the expected and actual results for small subsets of data. A major problem here is that when conducting such evaluations over publicly available research datasets, such as the MDC or GeoLife datasets used in this paper, or alternatives such as Yonsei, there is no mechanism for contacting  users to have them verify assumptions. To overcome this problem, we opt to use data collected ourselves for this part of the evaluation, as it affords us the ability to discuss with users exactly what activities they were conducting on a given day. Details of the data collected can be found in Section~\ref{sec:data}.

This section presents analyses on small amounts of manually labelled real-world data with the goal of using the constructed context trees to provide meaning to high-level behaviours, with the overall aim of identifying such behaviours from the tree. The data analysed spans 24 hours from the three users of the Warwick dataset, where annotations were added manually as accurately as possible, and in consultation with the users. The augmentation and filtering procedures were run over this data and, for each labelled time period, the 3 most common element tags were identified. This is shown in Figures~\ref{fig:gt:table1}-\ref{fig:gt:table3}. The aim here is not to label the time periods with the exact activity being performed, but rather to demonstrate that a meaningful relationship exists between the tags extracted and the true activity. 

In Figure~\ref{fig:gt:table1}, general labels are applied to the activities being performed, and a meaningful correlation between the tags extracted by the procedure and these labels is evident. Specific examples include the action of driving being labelled with the `highway' key, and taking the train with `railway'. Although the tags are not always perfect, they are indicative. For instance, when the individual was at home no residential building was identified, but an indication of the type of location was given by the tags `lit:yes' and `highway'. In the region where this data was collected, roads with street lighting typically signify residential areas. A similar relationship is shown in Figures~\ref{fig:gt:table2} and~\ref{fig:gt:table3}, with labels applied hierarchically and at lower granularities. While  not every item is labelled exactly, we believe this is a result of the data collection method. We used a data collection rate of one point per minute, meaning that several labelled activities consist of only 1 or 2 trajectory points, leaving little information for the procedure to utilise. Similarly, the land usage dataset contains a vast amount of information, but can be limited in parts. An example of this is that the pub which was visited at 17:25 (Figure~\ref{fig:gt:table2}) is inside a larger building. The procedure is only capable of identifying that the building was occupied by the user, but there is no information pertaining to which element inside the building was being interacted with, and so the only available information is `building:yes'.

\begin{figure}[t]
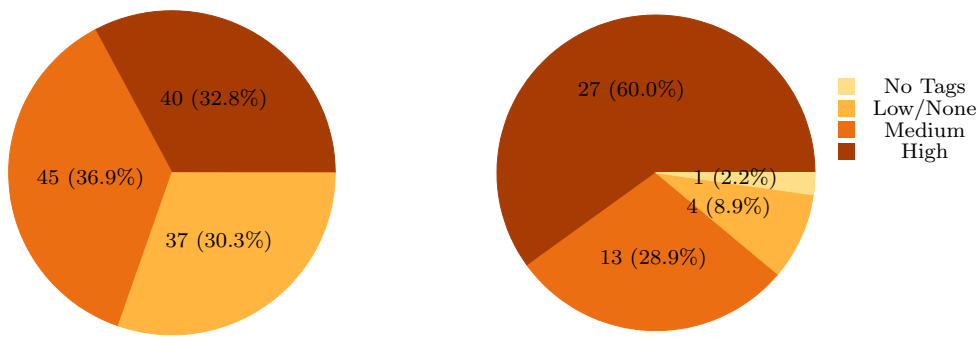

  \centering
  \footnotesize
  \begin{subfigure}[b]{0.4\textwidth}
    \footnotesize
    \input{figures/piecharts/16_a}
    \caption{Overall tag relevancy}\label{fig:eval:gt:weighted} 
  \end{subfigure}
  \begin{subfigure}[b]{0.58\textwidth}
    \footnotesize
    \input{figures/piecharts/16_b}
    \caption{Best tag relevancy per time period}\label{fig:eval:gt:firsts}
    \end{subfigure}
  \caption{Proportion of relevant tags.}\label{fig:eval:gt:both}
\end{figure}

To quantitatively explore how well the procedure worked over these examples, each tag extracted is scored based on the relevancy to the label using three classifications: \emph{high}, \emph{medium}, \emph{low/none}. These scores are manually assigned and shown in Table~\ref{tab:tag_counts}, where a \emph{high} tag indicates that the label is very well correlated to the activity (e.g. `building:residential' to the activity `Home'), \emph{medium} indicates that there is some link (e.g. `surface:asphalt' to `Driving on a main road'), and \emph{low/none} being given to tags with little or no relationship to the activity (e.g. `highway:bus\_stop' to `Attending lecture'). Figure~\ref{fig:eval:gt:weighted} shows the proportion of tags assigned to each of these weightings, demonstrating that the procedure identified tags with \emph{high} or \emph{medium} relevancy 69.7\% of the time. We also consider the highest-ranked tag assigned to each labelled time period and the proportion of time periods represented by each tag score is shown in Figure~\ref{fig:eval:gt:firsts}. From these results, it is clear that while in the three examples, only 32.8\% of tags were awarded a \emph{high} relevancy score, 60.0\% of labels have at least one tag with such a score, and 88.9\% contain at least one tag with a score of \emph{high} or \emph{medium}. This indicates that while not all of the 3 tags per label were useful, in nearly all cases, at least one of them was. 

\begin{figure}[t]
   \centering
  \includegraphics[width=0.7\linewidth]{figures/label_table_1}
   \caption{Manually labelled data (in bold) compared against extracted element labels.}\label{fig:gt:table1}
\end{figure}

\begin{figure}[t]
   \centering
  \includegraphics[width=1\linewidth]{figures/label_table_2}
   \caption{Manually labelled data (in bold) compared against extracted element labels.}\label{fig:gt:table2}
\end{figure}

\begin{figure}[t]
   \centering
  \includegraphics[width=1\linewidth]{figures/label_table_3}
   \caption{Manually labelled data (in bold) compared against extracted element labels.}\label{fig:gt:table3}
\end{figure}

\begin{table}[p]
\centering
\caption{Summary of tags and frequency count for each type of interactions scored based on the relevancy of each tag (High, Medium and Low/None).}\label{tab:tag_counts}
\begin{tabular}{lllllll}
\textbf{Label}  & \textbf{Tag}                & \textbf{S} & \textbf{\#} & \textbf{Tag}                        & \textbf{S} & \textbf{\#} \\ \hline
Home            & landuse:residential        & H          & 2           & barrier:kissing\_gate              & L          & 1           \\
                & highway:residential        & H          & 2           & oneway:no                          & L          & 1           \\
                & building:residential       & H          & 1           & maxspeed:30                        & L          & 1           \\
                & building:garage            & M          & 1           & highway:primary                    & L          & 1           \\
                & lit:yes                    & M          & 1           & left\_county:nor...                & L          & 1           \\ \hline
Walking (res.)  & landuse:residential        & H          & 1           &                                     &            &             \\ \hline
Walking (shops) & amenity:parking            & M          & 1           &                                     &            &             \\ \hline
Walking (road)  & sidewalk:both              & H          & 2           & highway:bus\_stop                  & M          & 1           \\
                & highway:secondary          & H          & 1           & bicycle:yes                        & M          & 1           \\
                & oneway:yes                 & M          & 2           & ref:lmngtns                        & L          & 1           \\
                & lit:yes                    & M          & 2           & public\_transport:pay...           & L          & 1           \\
                & boundary:public...         & L          & 1           &                                     &            &             \\ \hline
Walking (park)  & leisure:park               & H          & 1           & waterway:river                     & M          & 1           \\
                & foot:yes                   & H          & 1           & barrier:gate                       & M          & 1           \\
                & barrier:kissing\_gate      & M          & 1           &                                     &            &             \\ \hline
Driving (res.)  & landuse:residential        & H          & 2           &                                     &            &             \\ \hline
Driving (road)  & highway:tertiary           & H          & 6           & maxspeed:60                        & M          & 3           \\
                & highway:primary            & H          & 2           & maxspeed:30                        & M          & 2           \\
                & highway:secondary          & H          & 1           & maxspeed:20                        & M          & 2           \\
                & oneway:yes                 & M          & 4           & amenity:university                 & L          & 2           \\
                & highway:bus\_stop          & M          & 3           & type:multipolygon                  & L          & 1           \\
                & surface:asphalt            & M          & 3           &                                     &            &             \\ \hline
Parking (uni)   & amenity:university         & M          & 1           & type:multipolygon                  & L          & 1           \\ \hline
Work (office)   & building:university        & H          & 2           & highway:footway                    & L          & 1           \\
                & building\_levels:4         & M          & 2           & highway:service                    & L          & 1           \\ \hline
Walking (uni)   & amenity:university         & H          & 4           & landuse:grass                      & M          & 1           \\
                & highway:crossing           & H          & 2           & type:multipolygon                  & L          & 4           \\ \hline
Eating (rest.)  & level:0                    & M          & 1           & area:yes                           & L          & 1           \\
                & level:1                    & M          & 1           & lit:yes                            & L          & 1           \\
                & building:yes               & M          & 1           & surface:asphalt                    & L          & 1           \\ \hline
Eating (pub)    & building:yes               & M          & 1           & area:yes                           & L          & 1           \\
                & level:0                    & M          & 1           &                                     &            &             \\ \hline
Work (library)  & amenity:library            & H          & 2           & type:multipolygon                  & L          & 2           \\
                & amenity:university         & M          & 2           &                                     &            &             \\ \hline
Work (lecture)  & surface:asphalt            & L          & 2           & highway:bus\_stop                  & L          & 1           \\
                & type:multipolygon          & L          & 1           & oneway:yes                         & L          & 1           \\
                & lit:yes                    & L          & 1           &                                     &            &             \\ \hline
Visiting friend & amenity:university         & M          & 1           & type:multipolygon                  & L          & 1           \\
                & building:yes               & M          & 1           &                                     &            &             \\ \hline
Petrol station  & operator:tesco             & H          & 1           & amenity:fuel                       & H          & 1           \\
                & opening\_hours:24/7        & H          & 1           &                                     &            &             \\ \hline
Union (uni)     & amenity:university         & M          & 1           & type:multipolygon                  & L          & 1           \\ \hline
Bar             & building:yes               & M          & 2           & oneway:yes                         & L          & 2           \\
                & surface:asphalt            & L          & 2           &                                     &            &             \\ \hline
Train           & electrified:rail           & H          & 2           & railway:rail                       & H          & 1           \\
                & gauge:1435                 & H          & 2           & frequency:0                        & L          & 1           \\ \hline
\end{tabular}
\end{table}
 
While this evaluation is limited in that it only considers 3 days worth of data from 3 different users, it provides an indication that the techniques discussed previously are extracting useful and correct elements. This is demonstrated by showing that there is a strong relationship between the tags identified by the system and the labels manually assigned to data as a partial ground truth. A complete ground truth is not possible in this domain, since the desirable properties of context trees will vary significantly based on their intended use, however we believe that this exploration goes some way to demonstrating the accuracy of the technique.

\FloatBarrier

\subsection{Context Trees}\label{sec:eval:trees}

\begin{figure}[t!]
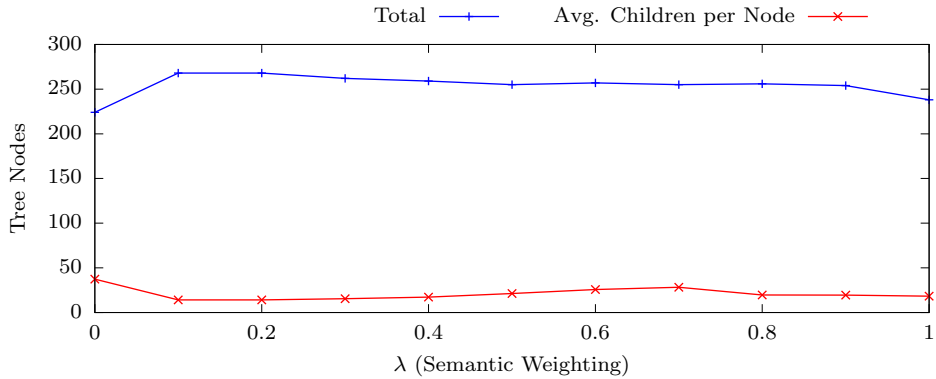

  \centering
  \footnotesize
  \begin{gnuplot}[terminal=epslatex, terminaloptions={size 5,2 font 8}]
    set key above
    set ylabel "Tree Nodes"
    set xlabel "$\\lambda$ (Semantic Weighting)"
    set yrange [0:]
    plot "data/clustering-6073-lambda.dat" u 1:2 with linespoints lt rgb "blue" lw 2 t 'Total', "data/clustering-6073-lambda.dat" u 1:3 with linespoints lt rgb "red" lw 2 t 'Avg. Children per Node'
  \end{gnuplot}
  \caption{Relationship between $\lambda$ and number of tree nodes.}\label{fig:clustering:lambda:total}
\end{figure} 

When constructing context trees from summarised data (Section~\ref{sec:clustering}), the only required parameter is $\lambda$, which specifies the weighting to be given to \textit{semantic similarity} as part of the \textit{Hybrid Contextual Distance} distance metric (Equation~\ref{eqn:hcd}). A weighting of 1 will construct a tree based only on the semantic similarity between node tags, and a weighting of 0 will construct a tree based only on the similarity of features, with any value in between using a combination of the two. The relationship between $\lambda$ and the number of nodes in a context tree is shown in Figure~\ref{fig:clustering:lambda:total} (generated using 24 hours of a single users' data, filtered with parameters $\delta = 1200$, $t = 0.8$, and $t_{max} = 1200$). While the number of nodes does not vary drastically with $\lambda$, the meaning behind the clusters does.

\begin{figure}[t!]
   \centering
  \includegraphics[width=0.6\linewidth]{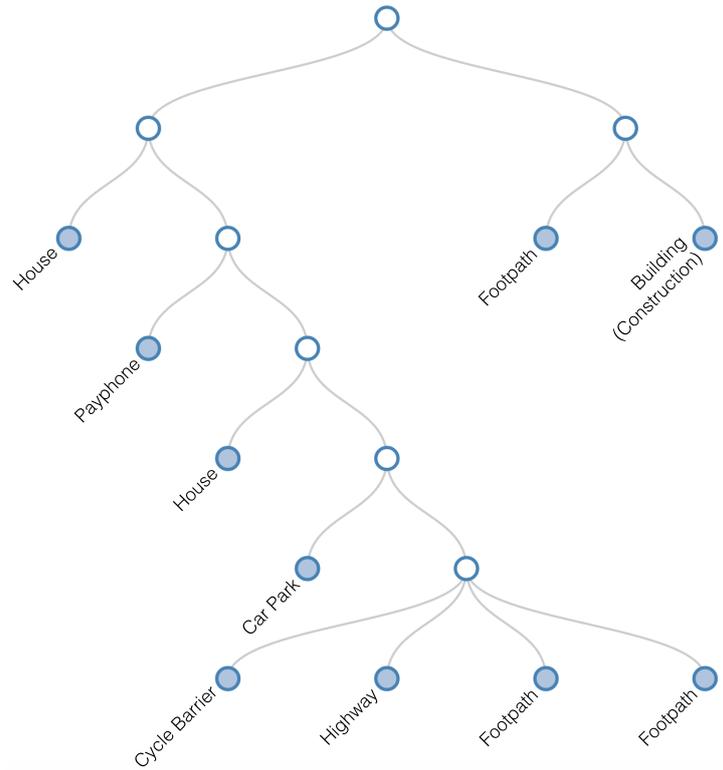}
   \caption{Example context tree: geographic clustering.}\label{fig:tree:geographic}
\end{figure}

\begin{figure}[t!]
   \centering
  \includegraphics[width=0.7\linewidth]{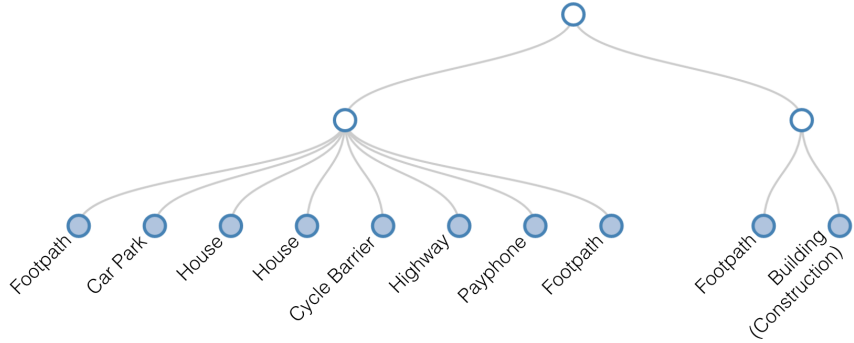}
   \caption{Example context tree: temporal clustering.}\label{fig:tree:temporal}
\end{figure}

\begin{figure}[t]
   \centering
	\includegraphics[width=0.8\linewidth]{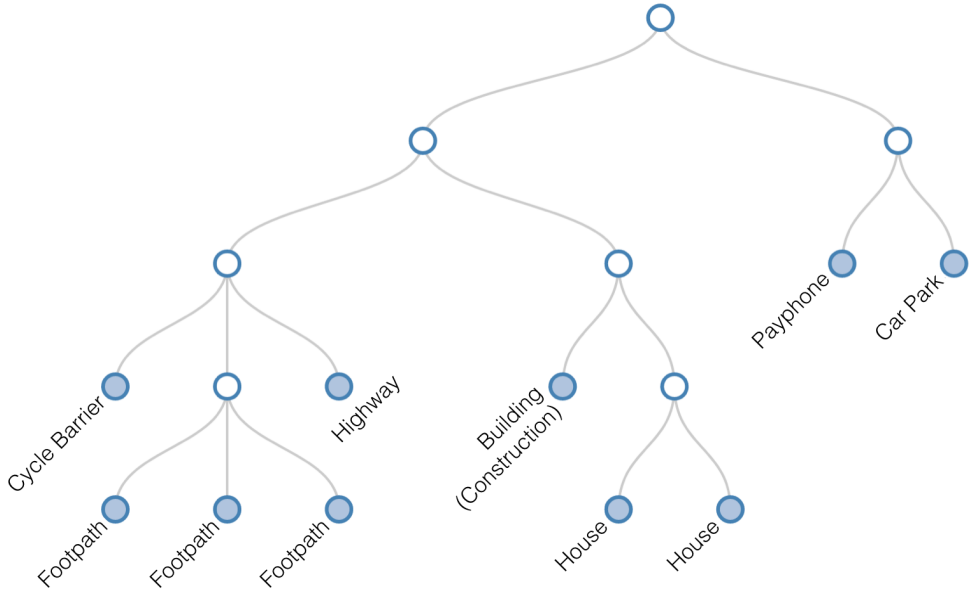}
   \caption{Example context tree: semantic clustering ($\lambda = 1$).}\label{fig:tree:semantic}
\end{figure}

\begin{figure}[t]
   \centering
	\includegraphics[width=0.8\linewidth]{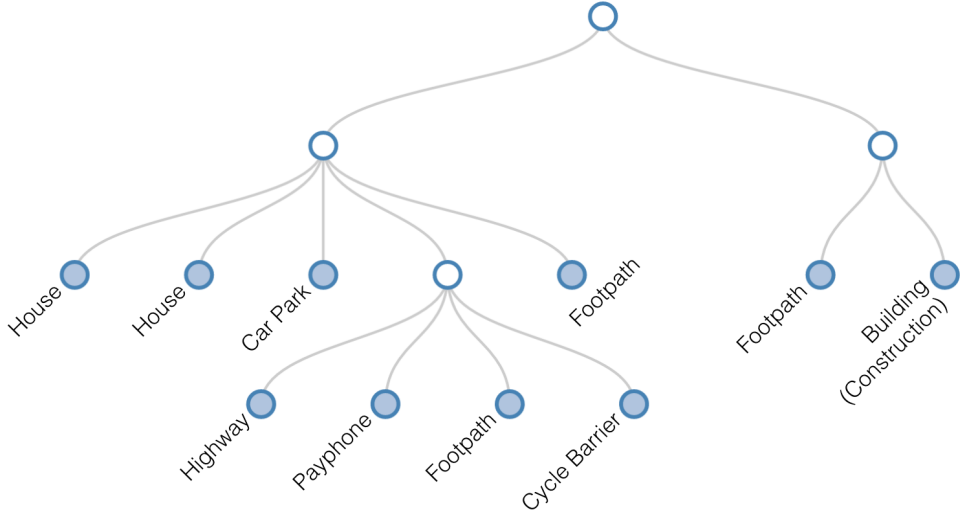}
   \caption{Example context tree: feature-based clustering ($\lambda = 0$).}\label{fig:tree:features}
\end{figure}

\begin{figure}[t]
   \centering
	\includegraphics[width=0.8\linewidth]{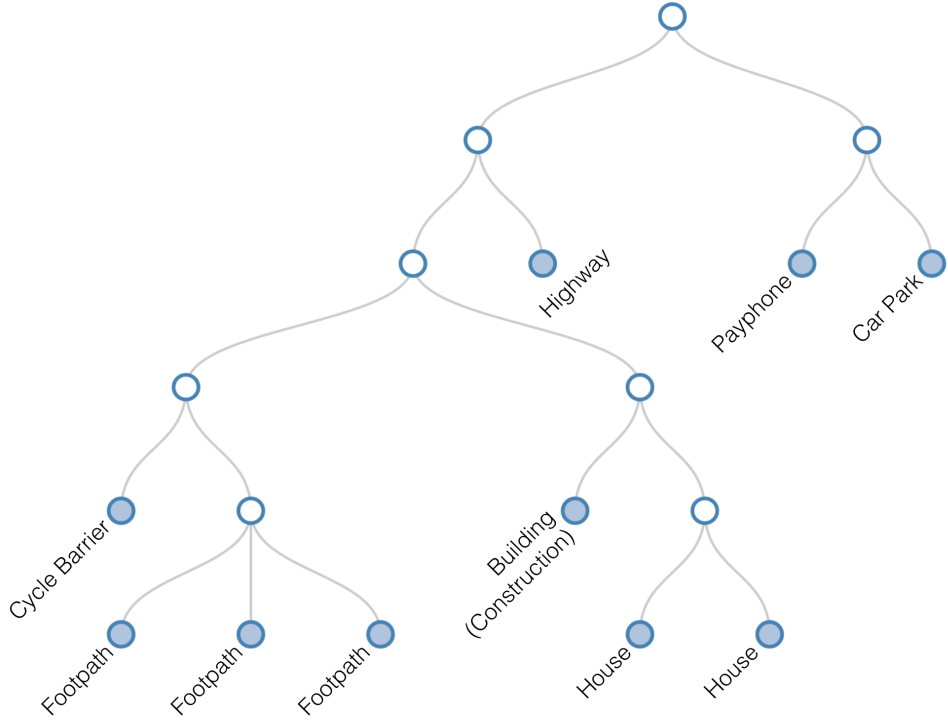}
   \caption{Example context tree: hybrid clustering ($\lambda = 0.6$).}\label{fig:tree:hybrid}
\end{figure}

Since our work on understanding context from trajectories augmented with land usage information is novel, there are no existing baseline methods or ground truth datasets to compare against. Instead, we take the closest method to a baseline that exists and compare the results against this. Figures~\ref{fig:tree:geographic} and~\ref{fig:tree:temporal} show the results of clustering context trees using na\"{i}ve distance metrics that consider only geographic distance between elements (Figure~\ref{fig:tree:geographic}) and temporal distance between interactions (Figure~\ref{fig:tree:temporal}). While these figures only show one small example, the results are representative of using such metrics in that the elements clustered together have no clear contextual relationship. This is in contrast to the context trees generated from the same data using the Hybrid Contextual Distance metric, along with different values of $\lambda$, as shown in Figures~\ref{fig:tree:semantic}--\ref{fig:tree:hybrid}.

In all of these examples, the element identifier has been manually replaced with a descriptive keyword to represent the element. Semantic clustering (Figure~\ref{fig:tree:semantic}) creates distinctive groups for buildings, footpaths and public amenities, as the elements in these groups are similar, while feature-based clustering (Figure~\ref{fig:tree:features}) creates groups that are less easily identifiable and relate to properties of the elements (e.g.\ the footpaths are not grouped because they were not encountered in the same journey, but rather were used at different times of the day). Finally, hybrid clustering (Figure~\ref{fig:tree:hybrid}) shows properties of both semantic and feature-based clustering where both the description of the element and properties of the interaction with the element are considered to create clusters. Selecting an appropriate value of $\lambda$ is application-specific.

These context trees provide only small examples of the differences between trees generated with na\"{i}ve distance metrics (Figures~\ref{fig:tree:geographic} and~\ref{fig:tree:temporal}) and those generated with the HCD metric (Figures~\ref{fig:tree:semantic}--\ref{fig:tree:hybrid}). In order to quantify such differences, and given knowledge of the data and how it was collected, we opt to make several assumptions of expected properties of generated context trees and explore the extent to which these expectations are violated with each distance metric. While this is of course a subjective evaluation, and the utility will vary based on the specific application the context tree is put to, it goes some way to providing an indicator of the utility of this approach in lieu of a ground truth. The assumptions made are:

\begin{enumerate}
\item Buildings should be grouped together unless they have very different uses (e.g. residential buildings should not be in the same group as office buildings).
\item Roads should be grouped together, with elements relating to roads grouped at a higher level (e.g. junctions).
\item Public amenities should be grouped together unless the interactions have very different properties.
\end{enumerate} 

These assumptions focus on the semantics of elements, but the features also need to be considered when exploring possible reasons for clusters being split. For instance, if a person visited many houses as part of their job, it would be reasonable to assume that these houses should be semantically close to the residence of the individual in the context tree, but not at exactly the same level. The usefulness of such assumptions will depend on the application, but it is possible to see that when aiming to characterise how a person has spent their time, it is beneficial to identify the times spent at residential buildings  separately to those spent at work. On the small example context trees shown in this section, geographic and temporal clustering (Figures~\ref{fig:tree:geographic} and~\ref{fig:tree:temporal}) violate all 3 assumptions. Semantic clustering (Figure~\ref{fig:tree:semantic}) best adheres to these assumptions, with the houses grouped at the same level and the building under construction close by in the next level up. Similarly, the footpaths are together with the cycle barrier, a related element, and highway one level up. Feature-based clustering (Figure~\ref{fig:tree:features}) has fewer valid assumptions than semantic clustering, as it only considers the interactions with the elements and not the elements themselves. Although the houses are together in a single cluster, they are also joined with the car park and footpath. Finally, hybrid clustering (Figure~\ref{fig:tree:hybrid}) is very similar to semantic clustering with the exception that the highway is no longer situated close to the footpaths, but is further up the context tree by itself. This still leaves 2 of the assumptions strictly adhered to, with 1 very close. A change that can be explained by the consideration of interaction features, where the highway has a different profile of interaction than the footpath and cycle barrier elements. Again, these are  small examples, however the trends present have been observed to be consistent across larger context trees.

\begin{figure}[t]
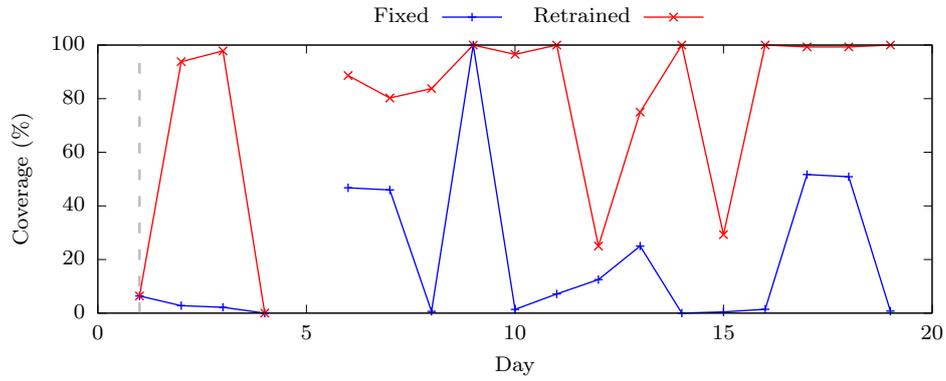
 
  \centering
  \footnotesize
  \begin{gnuplot}[terminal=epslatex, terminaloptions={size 5,2 font 8}]
    set datafile missing "?"
    set key above 
    set ylabel "Coverage ($\\
    set xlabel "Day"
    set arrow from 1,graph(0,0) to 1,graph(1,1) nohead dashtype 2 lc rgb 'grey' lw 4
    plot "data/clustering-6073-coverage.dat" using 1:($3) with linespoints lt rgb "blue" lw 2 t "Fixed", "data/clustering-6073-coverage.dat" using 1:($4) with linespoints lt rgb "red" lw 2 t "Retrained"
  \end{gnuplot}
  \caption{Land usage coverage with an initial training period of 24 hours (indicated by the dotted line).}\label{fig:res:filter:coverage}
\end{figure}

\begin{figure}[t]
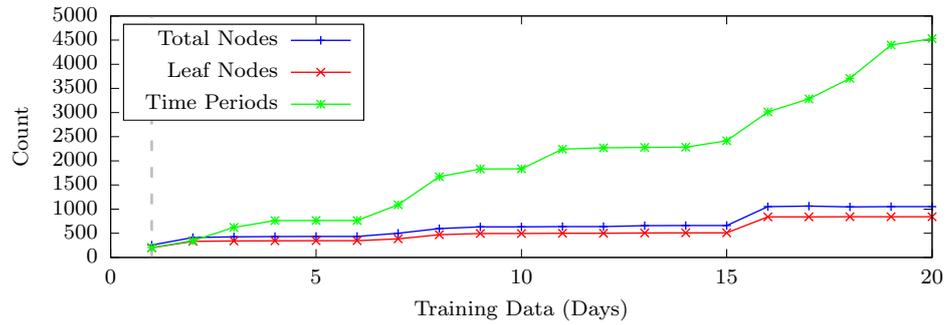
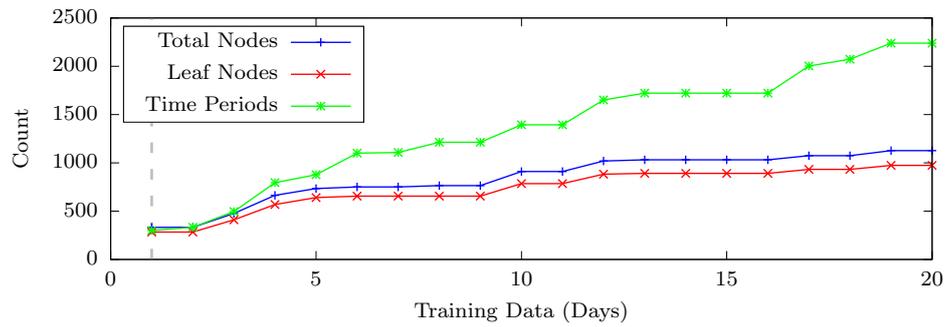

  \centering
  \footnotesize
  \begin{subfigure}[b]{1\textwidth}
  \centering
    \begin{gnuplot}[terminal=epslatex, terminaloptions={size 5,1.75 font 8}]
      set key left box opaque
      set key spacing 1.5
      set ylabel "Count"
      set xlabel "Training Data (Days)"
      set yrange [0:]
      set xrange [0:]
      set arrow from 1,graph(0,0) to 1,graph(1,1) nohead dashtype 2 lc rgb 'grey' lw 4
      plot "data/clustering-6073-rebuilding.dat" u ($1 / 24):2 with linespoints lt rgb "blue" lw 2 t 'Total Nodes', "data/clustering-6073-rebuilding.dat" u ($1 / 24):3 with linespoints lt rgb "red" lw 2 t 'Leaf Nodes', "data/clustering-6073-rebuilding.dat" u ($1 / 24):4 with linespoints lt rgb "green" lw 2 t 'Time Periods' #$
    \end{gnuplot}
    \caption{MDC user}\label{fig:clustering:nodecount:mdc}
  \end{subfigure}
  \begin{subfigure}[b]{1\textwidth}
  \centering
    \begin{gnuplot}[terminal=epslatex, terminaloptions={size 5,1.75 font 8}]
    set key left box opaque
    set key spacing 1.5
    set ylabel "Count"
    set xlabel "Training Data (Days)"
    set yrange [0:]
    set xrange [0:]
    set arrow from 1,graph(0,0) to 1,graph(1,1) nohead dashtype 2 lc rgb 'grey' lw 4
    plot "/Users/ali/Data/GeoLife/Graphs/threshold--1200--GEO_079.yml" u ($1 / 24):2 with linespoints lt rgb "blue" lw 2 t 'Total Nodes', "/Users/ali/Data/GeoLife/Graphs/threshold--1200--GEO_079.yml" u ($1 / 24):3 with linespoints lt rgb "red" lw 2 t 'Leaf Nodes', "/Users/ali/Data/GeoLife/Graphs/threshold--1200--GEO_079.yml" u ($1 / 24):4 with linespoints lt rgb "green" lw 2 t 'Time Periods' #$
    \end{gnuplot}
    \caption{GeoLife user}\label{fig:clustering:nodecount:geolife}
  \end{subfigure}
  \caption{Training data against number of tree nodes.}\label{fig:clustering:nodecount}
\end{figure}

With a better understanding of filtering, summarising and clustering, we turn our attention to exploring how data influences the properties of the generated context tree. Focusing on 21 days of data from a single user, Figure~\ref{fig:res:filter:coverage} shows repetition in data by using the first day as a set of \emph{training} data and calculating the coverage (i.e.\ the proportion of \emph{test} data present in the \emph{training} data) for each following day, shown by the blue \emph{Fixed} line. Additionally, the red \emph{Retrained} line shows the coverage when using all previous days (i.e.\ $0$ to $n-1$, where $n$ is the current day) as the \emph{training} set. The total number of nodes, number of leaf nodes, and total count of time periods for a context tree generated using the same data (where day $n$ shows a summary for a tree built using all data from days $0$ to $n$) are shown in Figure~\ref{fig:clustering:nodecount:mdc}. Please note that no data was recorded during day 5 for this sample user in the MDC dataset.

Figure~\ref{fig:res:filter:coverage} begins with a low coverage for both \emph{Fixed} and \emph{Retrained} lines, indicating that few elements encountered in day 1 were present in the training set (day 0). However, while the \emph{Fixed} score remains low for days 2--4, the \emph{Retrained} score approaches 100\%. In this instance, this is indicative of the user visiting elements that they did not encounter in the initial training day (day 0), but that they did encounter during subsequent days, as the \emph{Retrained} line includes all previous days as training data. The figure shows similar results for the remaining test days, where during day 9 the user visited only locations visited during day 0 and during days 9, 11 and 16--20 the user encountered no new elements as the score for \emph{Retrained} is at 100\%. Figure~\ref{fig:clustering:nodecount:mdc} shows how these properties relate to the size of context trees generated. The number of \emph{leaf nodes} is the number of unique elements and the number of \emph{time periods} is a count of the total number of (non-unique) elements encountered. That is, if the user encountered the same element 3 times, or 3 different elements, both would count as 3 time periods. At day 1 the number of time periods is roughly the same as the number of leaf nodes, indicating that all elements were encountered approximately once. As time goes by, more elements are encountered, but a large number of existing elements are revisited, demonstrated by the disproportionate rise in the number of time periods. This indicates that over a short period, where the user likely remained within a single region, the size of the tree does not increase significantly as additional data is added. However, considering trees over larger time periods will not have the same property as the user will likely visit new regions with entirely new leaf nodes. Figure~\ref{fig:clustering:nodecount:geolife} shows a similar graph as Figure~\ref{fig:clustering:nodecount:mdc}, however it was generated using data from a user of the GeoLife dataset instead of the MDC dataset. As is evidenced by the figures, the procedure extracts similar trends in users from each dataset.

This section has characterised the outputs and properties of the context tree generation procedure presented in Section~\ref{sec:overview}. While the concept of a  ground truth for this work is not applicable, and existing approaches for comparison are lacking, through the provision of multiple small examples and a discussion of general trends we have demonstrated the applicability of the approach presented in this paper to the task of identifying similar contexts and storing such information in a hierarchical data structure.
\FloatBarrier
\section{Context Tree Pruning}\label{sec:pruning}

\begin{figure}[t] 
  \centering
  \includegraphics[width=0.5\linewidth]{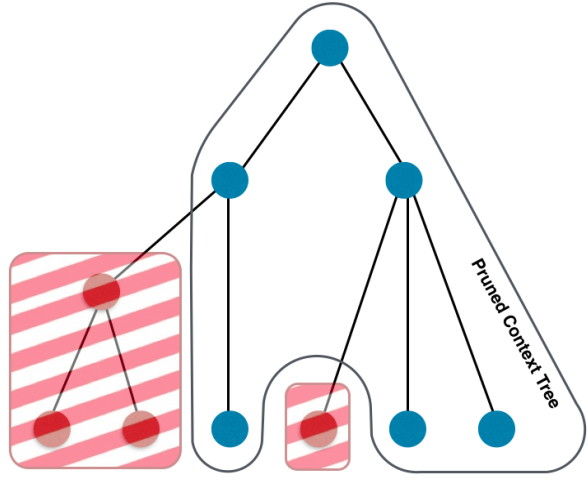}
  \caption{An example of a pruned context tree (with removed nodes crossed through).}\label{fig:diagram_pruned_tree}
\end{figure}

Storing context trees in their entirety maintains the maximum amount of information, however there are applications where reducing the size of a tree may be desirable. Memory-constrained devices, for example, may be better able to make use of a reduced size context tree as this would require lower storage requirements, and also enable quicker search due to the reduced number of nodes. Furthermore, reducing the size of context trees may have application-specific benefits, such as preventing overfitting when learning prediction models. In both of these cases, it is desirable to \emph{prune} the tree to reduce the amount of data stored while maintaining as much information as possible. This section presents a method for such pruning, that although requires additional processing to select nodes eligible to be removed, results in smaller context trees that require less memory to store and fewer operations to search. A representation of a pruned context tree can be seen in Figure~\ref{fig:diagram_pruned_tree}.

\subsection{Pruning Criteria}\label{sec:pruning:criteria}

Pruning is performed depth-first, evaluating each cluster to determine whether the additional overhead of storing the node is outweighed by the utility it affords. Each cluster is considered using the \emph{null hypothesis}, and the hypothesis rejected when the utility of storing the cluster is above a threshold. Any cluster for which we are unable to reject the hypothesis is pruned, and its parent is marked as eligible for pruning. As metrics do not already exist for this task, we adapt existing metrics used in related domains for the purpose of context tree pruning.

\subsubsection{Storage Cost}\label{sec:pruning:cost}

Clusters are scored according to two metrics: their storage cost and their utility. To determine the cost of storing a cluster, it is important to understand how clusters are built up in a context tree (described in Section~\ref{sec:clustering:merging}). When merging two clusters together to form a parent cluster, the aspects that belong to each cluster are considered in turn; specifically the \emph{tags}, \emph{times} and \emph{coordinate sets}. Sets of tags are combined from the child clusters by taking their union, while times and coordinate sets are merged in such a way that overlapping components are combined into single elements, and thus through the combining of child clusters into a parent cluster, information has been removed. The cost of storing an additional node therefore is the cost of storing the individual components (e.g.\ time range) that are present in a child, but not present in the same form in its parent. Assuming uniform cost for each component:
\begin{multline}\label{eqn:cost}
\mathit{Cost}(C|P) = \\
\xi + |C_{times} \setminus P_{times}| + |C_{coordsets} \setminus P_{coordsets}| + |\cup_{s \in C_{coordsets}}{s} \setminus \cup_{s \in P_{coordsets}}{s}|
\end{multline}

\noindent Where $\xi > 0$ is a small, manually selected, penalty that represents the overhead of storing each cluster, $C_{times}$ is the set of time ranges that are associated with cluster $C$ and $C_{coordsets}$ is the set of coordinate sets associated with cluster $C$. Remembering that the coordinate sets belonging to a cluster themselves contain sets of points (i.e.\ $C_{coordsets}$ = $\{\{p_{1:1},p_{1:2},p_{1:3}, ...\},\{p_{2:1},p_{2:2},p_{2:3}, ...\}, ...\}$), $\cup_{s \in C_{coordsets}}{s}$ is taken to be the set of all points associated with any coordinate set that belongs to cluster $C$. Having $\xi$ as non-zero represents that there is always a (small) cost associated with each cluster. Equation~\ref{eqn:cost} will need tuning based on the specific application to better represent the true cost of storing a node, but it provides a basic foundation.

\subsubsection{Cluster Utility}\label{sec:pruning:utility}

Determining the utility of a cluster is difficult and is dependent on the specific use of the context tree. For this reason, any application of the approach will  need to consider the goal of pruning and use this to inform the measurement of the utility afforded by a specific cluster. We adopt a general approach that can be tailored to specific needs by providing a measure of the information lost if the parent of a cluster were used to represent the child, similar in idea to the \emph{Kullback-Leibler divergence} used to measure the difference between probability distributions. As parents contain a superset of the children, we consider the utility of a child cluster ($C$) given its parent ($P$) to be the proportion of information present in the parent that is not covered by the child, where the measure of information must consider the attributes (i.e.\ tags, times, and coordinate sets) present in each cluster:
\begin{equation}\label{eqn:information}
\textstyle
\mathit{Information(C)} = \sum_{t \in C_{times}}duration(t) + \sum_{s \in C_{coordsets}}{area(s)} + |C_{tags}|
\end{equation}

\noindent Providing even weighting to the different elements for the measure of utility:
\begin{equation}
\mathit{Utility}(C|P) = 1 - \left(\frac{1}{3}\frac{\sum_{t \in C_{times}}duration(t)}{\sum_{t \in P_{times}}duration(t)} + \frac{1}{3}\frac{\sum_{s \in C_{coordsets}}{area(s)}}{\sum_{s \in P_{coordsets}}{area(s)}} + \frac{1}{3}\frac{|C_{tags}|}{|P_{tags}|}\right)
\end{equation}

\noindent Specifically, this metric considers the proportion of time, area and tags covered by the child with respect to the parent, and holds true to the aims of such a metric to produce a score of 0 if the parent and child contain identical information and a score approaching 1 if the child only represents a fraction of the parent. 

\subsubsection{Cost-Benefit Score}

The \emph{cost-benefit score} of a cluster is taken to be the utility of the cluster divided by the storage cost:
\begin{equation}
\mathit{CostBenefitScore}(C|P) = \frac{\mathit{Utility}(C|P)}{\mathit{Cost}(C|P)}
\end{equation}

\noindent While utility is normalised between 0 and 1 as it represents the proportion of the parent that is not covered by the child, cost only has a minimum bound of $\xi$ (Section~\ref{sec:pruning:cost}), where $\xi > 0$. Depending upon the application, it may be desirable to also normalise cost relative to the current context tree. Using this metric on nodes depth-first, pruning should occur for any cluster $C$ with parent $P$ and $\mathit{CostBenefitScore}(C|P) < \theta$, where $\theta$ is the \textit{pruning threshold} and $C$ has no unpruned children.

\subsection{Pruning Evaluation}

Pruning requires a pre-built context tree and two parameters, namely $\theta$ and $\xi$, where $\theta$ provides a threshold for pruning, and $\xi$ is a penalty associated with every node when calculating its \emph{storage cost}.

\begin{figure}[t]
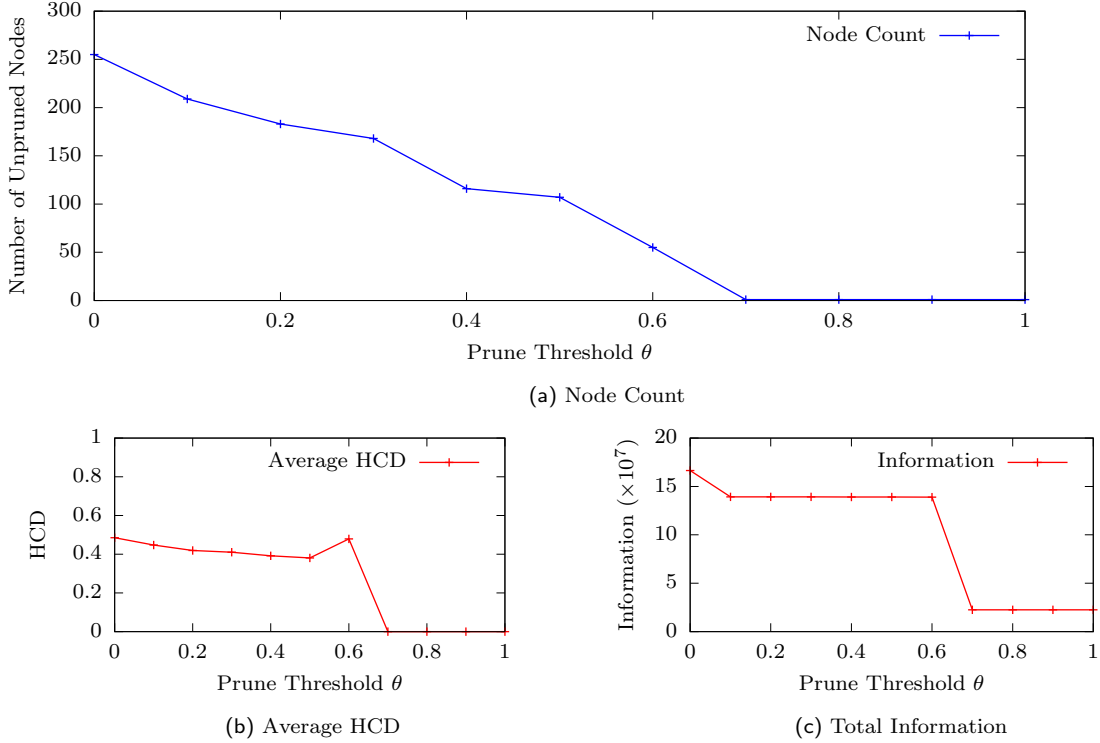

  \centering
  \footnotesize
  \begin{subfigure}[b]{1\textwidth}
  \begin{gnuplot}[terminal=epslatex, terminaloptions={size 5.5,2 font 8}]
    set ylabel "Number of Unpruned Nodes"
    set xlabel "Prune Threshold $\\theta$"
    set yrange [0:]
    set key off
    set key spacing 1.5
    plot "data/pruning-6073-theta.dat" using 1:3 with linespoints lt rgb "blue" lw 2 t "Node Count"
  \end{gnuplot}
  \caption{Node Count}\label{fig:pruning:theta:count}
  \end{subfigure}
  \begin{subfigure}[b]{0.48\textwidth}
  \begin{gnuplot}[terminal=epslatex, terminaloptions={size 2.7,1.5 font 8}]
    set ylabel "HCD"
    set xlabel "Prune Threshold $\\theta$"
    set yrange [0:1]
    set key off
    set key spacing 1.5
    plot "data/pruning-6073-theta.dat" using 1:5 with linespoints lt rgb "red" lw 2 t "Average HCD"
  \end{gnuplot}
  \caption{Average HCD}\label{fig:pruning:theta:hcd}
  \end{subfigure}
  \begin{subfigure}[b]{0.48\textwidth}
  \begin{gnuplot}[terminal=epslatex, terminaloptions={size 2.7,1.5 font 8}]
      set ylabel "Information ($\\times 10^{7}$)"
      set xlabel "Prune Threshold $\\theta$"
      set yrange [0:20]
      set key off
      set key spacing 1.5
      plot "data/pruning-6073-theta.dat" using 1:($6/1000000) with linespoints lt rgb "red" lw 2 t "Information" #$
  \end{gnuplot}
  \caption{Total Information}\label{fig:pruning:theta:info}
  \end{subfigure}
  \caption{Effect of $\theta$ on number of nodes in a sample context tree ($\lambda=0.5, \xi=1$).}\label{fig:pruning:theta}
\end{figure}

Figure~\ref{fig:pruning:theta} shows the effect of varying $\theta$ when pruning a context tree generated from the same data and parameters as those used in Figure~\ref{fig:clustering:lambda:total}, with $\lambda = 0.5$ and $\xi = 1$. From this figure it is possible to see that the number of nodes in a context tree can be drastically reduced while maintaining the majority of the information. Selecting $\theta = 0.6$, the resultant pruned context tree contains approximately 20\% of the nodes present in the unpruned tree, but maintains almost 70\% of the useful information. While the process to prune the context tree adds in additional complexity, the resultant tree is considerably more compact and thus applications that require storing or searching the tree will have significantly lower overhead.

\begin{figure}[t]
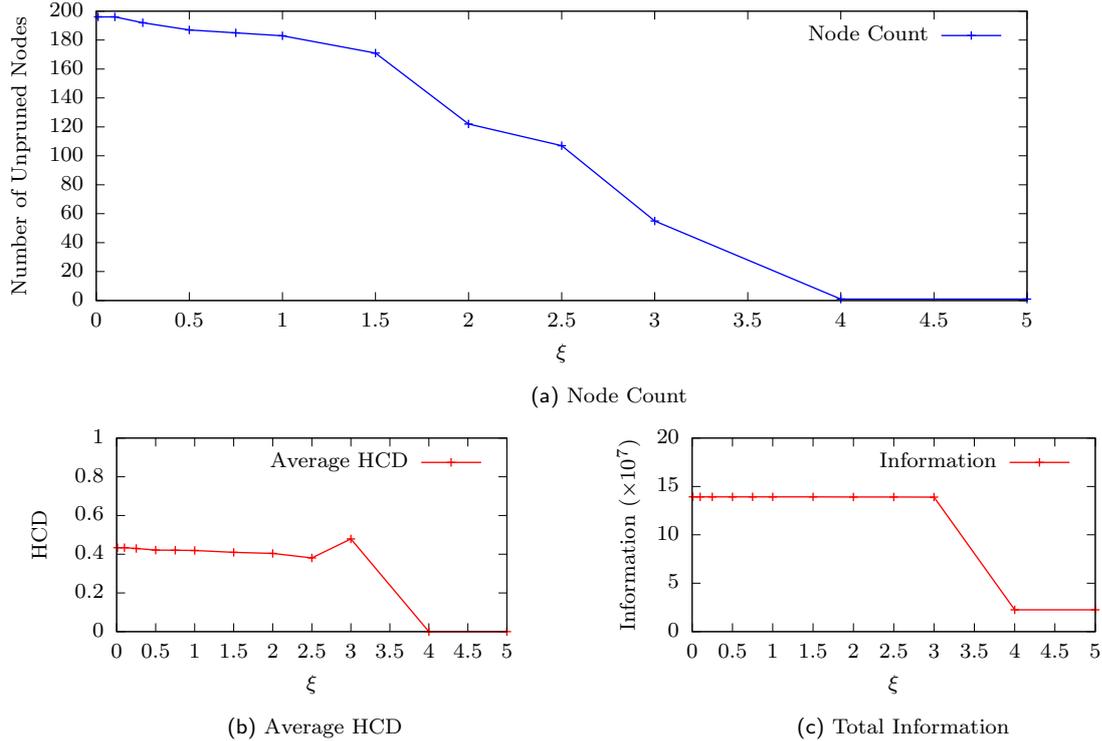

  \centering
  \footnotesize
  \begin{subfigure}[b]{1\textwidth}
  \begin{gnuplot}[terminal=epslatex, terminaloptions={size 5.5,2 font 8}]
    set ylabel "Number of Unpruned Nodes"
    set xlabel "$\\xi$"
    set yrange [0:]
    set key off
    set key spacing 1.5
    plot "data/pruning-6073-xi.dat" using 1:3 with linespoints lt rgb "blue" lw 2 t "Node Count"
  \end{gnuplot}
  \caption{Node Count}\label{fig:pruning:xi:count}
  \end{subfigure}
  \begin{subfigure}[b]{0.48\textwidth}
  \begin{gnuplot}[terminal=epslatex, terminaloptions={size 2.7,1.5 font 8}]
    set ylabel "HCD"
    set xlabel "$\\xi$"
    set yrange [0:1]
    set key off
    set key spacing 1.5
    plot "data/pruning-6073-xi.dat" using 1:5 with linespoints lt rgb "red" lw 2 t "Average HCD"
  \end{gnuplot}
  \caption{Average HCD}\label{fig:pruning:xi:hcd}
  \end{subfigure}
  \begin{subfigure}[b]{0.48\textwidth}
  \begin{gnuplot}[terminal=epslatex, terminaloptions={size 2.7,1.5 font 8}]
    set ylabel "Information ($\\times 10^{7}$)"
    set xlabel "$\\xi$"
    set yrange [0:20]
    set key off
    set key spacing 1.5
    plot "data/pruning-6073-xi.dat" using 1:($6/1000000) with linespoints lt rgb "red" lw 2 t "Information" #$
  \end{gnuplot}
  \caption{Total Information}\label{fig:pruning:xi:info}
  \end{subfigure}
  \caption{Effect of $\xi$ on number of nodes in a sample context tree ($\lambda=0.5, theta=0.2$).}\label{fig:pruning:xi}
\end{figure}

Using the same data again, but this time holding $\theta = 0.2$, Figure~\ref{fig:pruning:xi} shows the effect of changing $\xi$ on the number of unpruned nodes, average HCD and information. Increasing either $\theta$ or $\xi$ reduces the number of nodes left after pruning (Figures~\ref{fig:pruning:theta:count} and~\ref{fig:pruning:xi:count}), as increasing $\theta$ specifies a higher threshold required to maintain a node, and increasing $\xi$ assigns a higher cost to each node, making it less likely to exceed the threshold. The results also demonstrate that as more nodes are pruned from the context tree, the average distance of the remaining nodes becomes smaller (i.e. they become more similar, Figures~\ref{fig:pruning:theta:hcd} and~\ref{fig:pruning:xi:hcd}). Finally, Figures~\ref{fig:pruning:theta:info} and~\ref{fig:pruning:xi:info} demonstrate that although pruning does reduce the total information in the tree, it does so gradually until the number of unpruned nodes approaches 0, under the definition of information presented in Equation~\ref{eqn:information}. This helps to demonstrate the effectiveness of pruning as the number of nodes in the tree can be drastically reduced, but the amount of information remains high.

\begin{figure}[p]
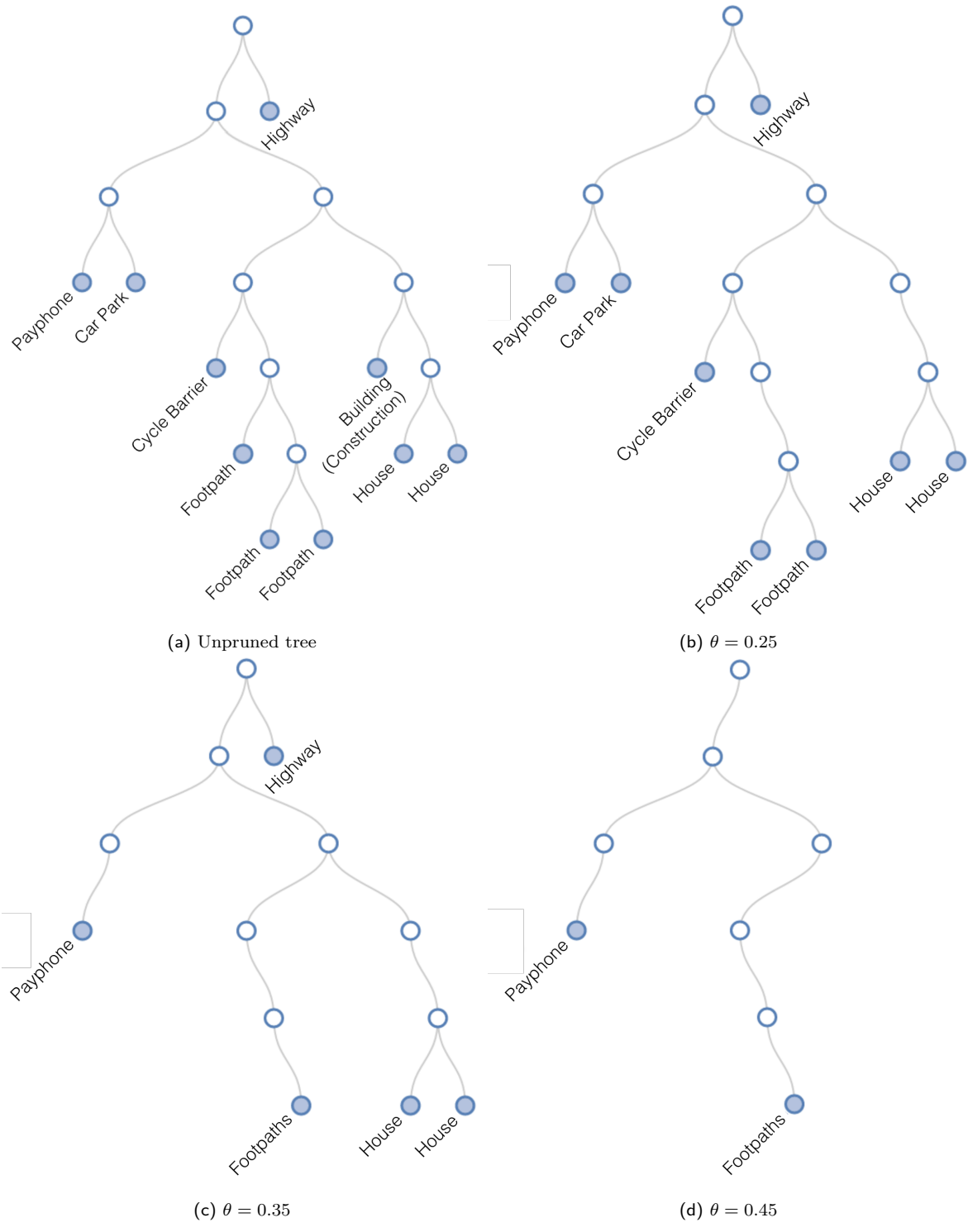

  \centering
  \footnotesize
  \begin{subfigure}[b]{0.48\textwidth}
    \includegraphics[width=1\linewidth]{figures/trees/pruning_6073_unpruned} 
    \caption{Unpruned tree}\label{fig:tree:pruned:unpruned}
  \end{subfigure}
  \begin{subfigure}[b]{0.48\textwidth}
    \includegraphics[width=1\linewidth]{figures/trees/pruning_6073_0-25_1-5-single} 
    \caption{$\theta = 0.25$}\label{fig:tree:pruned:0-25:1-5}
  \end{subfigure}
  \begin{subfigure}[b]{0.48\textwidth}
    \includegraphics[width=1\linewidth]{figures/trees/pruning_6073_0-35_1-5-single}
    \caption{$\theta = 0.35$}\label{fig:tree:pruned:0-35:1-5}
  \end{subfigure}
  \begin{subfigure}[b]{0.48\textwidth}
    \includegraphics[width=0.75\linewidth]{figures/trees/pruning_6073_0-45_1-5-single}
    \caption{$\theta = 0.45$}\label{fig:tree:pruned:0-45:1-5}
  \end{subfigure}
  \caption{Context tree pruning for different values of $\theta$, with $\xi = 1.5$.}\label{fig:tree:pruning}
\end{figure}

Figure~\ref{fig:tree:pruning} shows how pruning affects trees generated from real-world data (using the same data and clustering as in Figure~\ref{fig:tree:hybrid}). With the lowest value of $\theta$ ($\theta=0.25$ shown in Figure~\ref{fig:tree:pruned:0-25:1-5}), only two leaf nodes have been pruned: one of the footpaths and one of the buildings. Increasing $\theta$ ($\theta=0.35$ shown in Figure~\ref{fig:tree:pruned:0-35:1-5}) causes more leaf nodes to be pruned, and a further increase ($\theta=0.45$ shown in Figure~\ref{fig:tree:pruned:0-45:1-5}) has the effect of pruning entire sub-trees, resulting in a much smaller and more compact tree. Although containing less information, such pruned trees provide benefits in resource-constrained applications where storing and processing an entire tree may be infeasible.
\section{Conclusion}\label{sec:conclusion}

This work has presented the \emph{context tree} hierarchical data structure that summarises user contexts at multiple scales. In addition to this, we proposed a method for constructing context trees from geospatial trajectories and land usage datasets. The context tree is a novel data structure that provides rapid access to summary information about a user's interactions with their environment, and thus provides a foundation for further analysis, understanding and modelling of the behaviours of individuals and groups. Furthermore, this work has presented an analysis of both context trees and the associated generation procedure, using real-world data and a partial ground truth, alongside a proposed method of pruning context trees to reduce their size, thus requiring less processing and memory for further applications. The data employed for evaluation came from the publicly available MDC~\cite{Kiukkonen:2010vm,Laurila:2012vk} and GeoLife~\cite{Zheng:2008ku,Zheng:2009td,Zheng:2010uc} datasets, consisting of GPS trajectories from real individuals, in addition to data collected ourselves.

Constructing context trees begins with processing of land usage data in a manner that considers both the extraction of relevant land usage information and filtering to remove noise, in addition to providing a novel technique for clustering related land usage elements to expose contexts by considering both properties of the real-world entities that the user interacted with, and  properties of the interaction itself (e.g.\ the time and duration). These processes are combined with an agglomerative hierarchical clustering technique to generate the context tree.

By summarising contexts into a single data structure, it becomes easier to detect changes in routine through anomaly identification, identify similarities and differences between users to spot those with commonalities such as similar jobs or habits, and predict users' future actions. These areas are proving to be increasingly important to the provision of tailored and useful services both on individual and societal scales. Future work will expand existing techniques applied to locations and contexts by increasing their applicability to context trees. For example, expanding location prediction to operate over contexts such as those identified through contextual clustering would provide the ability to predict not only where a user is likely to be going, but also properties of the interaction, such as when and for how long. Furthermore, predictions need not relate to specific locations or entities, but rather to contexts and thus it would become possible to predict that a user will go to, for example, a building with certain properties without the need to identify exactly which building will be the target.

\section*{Acknowledgements}
Portions of the research in this paper used the MDC Database made available by Idiap Research Institute, Switzerland and owned by Nokia~\cite{Kiukkonen:2010vm,Laurila:2012vk}.

\FloatBarrier
\bibliographystyle{WarwickBibliography}
\bibliography{bibliography} 

\end{document}